\documentclass[sn-basic]{sn-jnl}

\usepackage[font=normal]{caption}
\usepackage{lineno,hyperref}
\usepackage{amsmath}
\usepackage{amstext}
\usepackage{adjustbox}
\usepackage{txfonts}
\usepackage{url}
\usepackage{relsize}
\usepackage{color}
\usepackage{bm}

\jyear{2022}%

\def\lsim{\,\lower2truept\hbox{${< \atop\hbox{\raise4truept\hbox{$\sim$}}}$}\,}
\def\gsim{\,\lower2truept\hbox{${> \atop\hbox{\raise4truept\hbox{$\sim$}}}$}\,}

\theoremstyle{thmstyleone}%
\theoremstyle{thmstyletwo}%

\theoremstyle{thmstylethree}%

\begin{document}

\title[Seafloor horizontal displacement from cGPS and compass data]{On the seafloor horizontal displacement from cGPS and compass data in the Campi Flegrei caldera}

\author*[1]{\fnm{Tiziana} \sur{Trombetti}}\email{trombetti@ira.inaf.it}

\author[1,2,3]{\fnm{Carlo} \sur{Burigana}}

\author[4,5]{\fnm{Prospero} \sur{De Martino}}

\author[4]{\fnm{Sergio} \sur{Guardato}}

\author[4]{\fnm{Giovanni} \sur{Macedonio}}

\author[4]{\fnm{Giovanni} \sur{Iannaccone}}

\author[1,6,7]{\fnm{Francesco} \sur{Chierici}}

\affil*[1]{\orgdiv{Istituto di Radioastronomia}, \orgname{INAF}, \orgaddress{\street{Via Piero Gobetti 101}, \city{Bologna}, \postcode{40129}, \state{Bologna}, \country{Italy}}}

\affil[2]{\orgdiv{Dipartimento di Fisica e Scienze della Terra}, \orgname{Universit\`a di Ferrara}, \orgaddress{\street{Via Giuseppe Saragat 1}, \city{Ferrara}, \postcode{44122}, \state{Ferrara}, \country{Italy}}}

\affil[3]{\orgdiv{Sezione di Bologna}, \orgname{INFN}, \orgaddress{\street{Via Irnerio 46}, \city{Bologna}, \postcode{40127}, \state{Bologna}, \country{Italy}}}

\affil[4]{\orgdiv{Osservatorio Vesuviano}, \orgname{Istituto Nazionale di Geofisica e Vulcanologia}, \orgaddress{\street{Via Diocleziano, 328}, \city{Napoli}, \postcode{80124}, \state{Napoli}, \country{Italy}}}

\affil[5]{\orgdiv{Istituto per il Rilevamento Elettromagnetico dell'Ambiente}, \orgname{Consiglio Nazionale delle Ricerche (IREA-CNR)}, \orgaddress{\street{Via Diocleziano, 328}, \city{Napoli}, \postcode{80124}, \state{Napoli}, \country{Italy}}}

\affil[6]{\orgdiv{Istituto di Scienze Marine}, \orgname{CNR}, \orgaddress{\street{Via Piero Gobetti 101}, \city{Bologna}, \postcode{40129}, \state{Bologna}, \country{Italy}}}

\affil[7]{\orgdiv{Sezione Roma 2}, \orgname{Istituto Nazionale di Geofisica e Vulcanologia}, \orgaddress{\street{Via di Vigna Murata 605}, \city{Roma}, \postcode{00143}, \state{Roma}, \country{Italy}}}

\abstract{Seafloor deformation monitoring is now routinely performed in the marine sector of the Campi Flegrei volcanic area (Southern Italy).
The MEDUSA infrastructure is formed by four buoys deployed at a water depth ranging from 40 to 96 m, and equipped with cGPS receivers, accelerometers and magnetic compasses to monitor the buoy status and
a seafloor module with a bottom pressure recorder and other onboard instruments. 
The analysis of the time series data acquired by the MEDUSA monitoring infrastructure system allows to study the seafloor deformation in the Campi Flegrei caldera with geodetic accuracy.
In a previous work we show that the time series acquired by the Campi Flegrei cGPS onland network and MEDUSA over the period 2017-2020 are in good agreement with the ground deformation field predicted by a Mogi model which is widely used to describe the observed deformation of  an active volcano in terms of magma intrusion.
Only for one of the buoys, CFBA (A), the data differ significantly from the model prediction, at a level of $\simeq$\,6.9\,$\sigma$ and of $\simeq$\,23.7\,$\sigma$ for the seafloor horizontal speed and direction, respectively. For this reason, we devised a new method to reconstruct the horizontal sea bottom displacement considering in the analysis both cGPS and compass data. 
The method, applied to the CFBA buoy measurements and validated also on the CFBC (C) buoy, uses compass data to correct cGPS positions accounting for the pole inclination. 
Including also systematic errors, the internal consistency, always within $\sim$\,3\,$\sigma$ for the speed and $\sim$\,2\,$\sigma$ for the angle, between the results derived for different maximum inclinations of the buoy pole (up to 3.5$\,^{\circ}$)
indicates that the method allows to significantly reduce the impact of the pole inclination which, if not properly taken into account, can alter the estimation of the horizontal seafloor deformation. In particular, we find a good convergence of the retrieved velocity and deformation angle as we include in the analysis data from increasing values of the buoy pole inclination.
Taking the result derived assuming the maximum allowed cutoff and accounting for statistical and systematic errors, we found a speed $\varv$ = (3.521 $\pm$ 0.039 (\textit{stat}) $\pm$ 0.352 (\textit{syst})) cm/yr and a deformation direction angle $\alpha$ = (-115.159 $\pm$ 0.670 (\textit{stat}) $\pm$ 7.630 (\textit{syst}))\,$^{\circ}$ (statistical errors at 1\,$\sigma$ quoted from the rms of their values, main systematic errors added linearly).
The relative impact of the main potential systematic (statistical) effects increases (decreases) with the cutoff. Our analysis provides a horizontal speed consistent with the model at a level of $\simeq$\,5.2\,$\sigma$ (\textit{stat} only) or of $\simeq$\,0.5\,$\sigma$ (\textit{stat} and \textit{syst} added linearly), and a deformation angle consistent with the model at $\simeq$\,4.3\,$\sigma$ level (\textit{stat} only) or at $\simeq$\,0.3\,$\sigma$ level (\textit{stat} and \textit{syst} added linearly). Correspondingly, the module of the vectorial difference between the velocity retrieved from the data and the velocity of the adopted Mogi model diminishes by a factor of $\simeq$ 7.65 $\pm$ 1.23 (\textit{stat}) or $\pm$ 5.78 (\textit{stat}+\textit{syst}) with respect to the previous work. A list of potential improvements to be implemented in the system and instruments is also discussed.}


\maketitle

\section{Introduction}\label{intro}

\begin{figure*}[h!]
\begin{center}
\includegraphics[width=11.5cm]{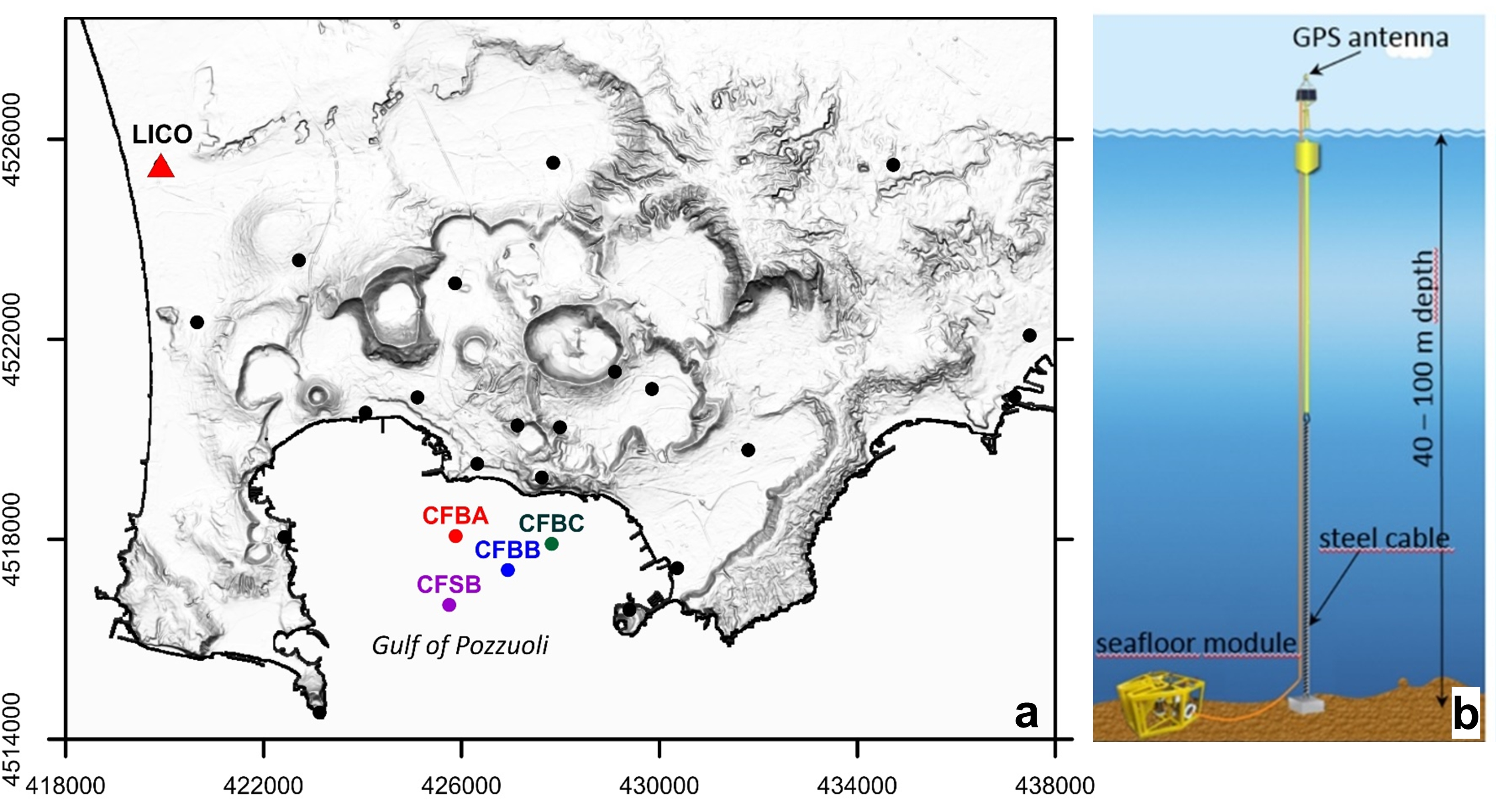}
\end{center}
\caption{Panel a: map of the Campi Flegrei area showing the locations of the onland cGPS stations and the four buoys (CFBA (A), CFBB (B), CFBC (C) and CFSB (CUMAS)) of the MEDUSA infrastructure in the Gulf of Pozzuoli. Panel b: scheme of a buoy showing the position of the GPS antenna.}
\label{fig:golfo}
\end{figure*}

MEDUSA is an innovative research infrastructure for monitoring shallow water seafloor displacement in Campi Flegrei volcanic area that consists of four anchored spar buoys installed in the Gulf of Pozzuoli at 1.1 to 2.4 km from the coast and water depths less than 100 m (see Fig. \ref{fig:golfo}). The buoys are equipped with geophysical and oceanographic instrumentation that transmit real time data to the INGV monitoring centre in Naples.

Among other instrumentation (see \citet{Iannaccone10}), cGPS receivers and digital compass sensors (\cite{Iannaccone09, Iannaccone18}) are installed
on the top of four buoys of MEDUSA, namely CFBA (A), CFBB (B), CFBC (C) and CFSB (CUMAS). 
The cGPS stations supply continuous position measurements along the North-South (NS), East-West (EW) and vertical (Up) directions. The data collected from the cGPS stations on the four MEDUSA buoys are processed in kinematic mode with the open source software RTKLIB ver. 2.4.2\footnote{http://www.rtklib.com}  to obtain positions every 30\,s. The cGPS station LICO, located at a distance of about 10 km from the caldera centre (red triangle in Fig. \ref{fig:golfo}), is the reference station in the data processing (a full description of the kinematic cGPS processing is reported in \cite{demartino14b}).

The compass sensors (see App. \ref{compass}), measuring the attitude of the buoy, provide three angles, roll, pitch and yaw (heading) of the buoy reference system, according to the Tait-Bryan convention (\cite{Henderson77}).

The measurements of both these instruments can be jointly used to recover the sea bottom deformation down to millimetre accuracy in presence of oscillations of the top of the buoy caused by marine and weather conditions (\cite{xie19}).

MEDUSA pole buoys, once deployed, behave like inverse pendulums. 
Their equilibrium positions, when unperturbed, stand along the vertical from the sea bottom. 
The most relevant deviations from the vertical equilibrium position are caused by environmental forces. 
Nonetheless undisturbed buoys, i.e. also in calm sea and absence of wind and currents, can be subject to very small vibration around their vertical position, due to mechanical resonances of the system.

The positions measured by the cGPS installed on the buoys result from a combination of the seafloor deformation and of the displacement of the pole from the equilibrium
due to environmental forcing. The latter can be much larger than the seafloor deformation and should be properly taken into account in the analysis and possibly accurately removed in order to recover the horizontal seafloor displacements. The recovery of the vertical component has already been described in \cite{demartino14b, demartino20}.

The Mogi model (\cite{mogi1958relations}) simulates a small spherical expansion source embedded in a homogeneous, isotropic, Poisson-solid half-space (\cite{Masterlark}), assuming that the crust is a semi-elastic medium and the occurring deformation is due to a source of pressure in the form of a spherical magma shape located at a certain depth (\cite{Sarsito_2019}). Thus, adopting a source location, depth and pressure change, this representation predicts the extent and magnitude of the resulting surface deformation pattern. We assume the convention that negative angles on the coordinate plane are angles that go in a clockwise direction and define the horizontal deformation angle, $\alpha$, caused by the seabed displacement in the caldera as the angle from the East direction.

A previous analysis of the horizontal time series acquired by MEDUSA buoys has been performed by \cite{demartino20} during the period from 1 July 2017 to 31 May 2020 using cGPS data only. Being the prototypical CUMAS station characterized by a very high noise level, it was not included in their horizontal component study.
Regarding the B and C buoys, the authors found a good agreement with the predictions of the Mogi model which best fits the displacement measured at the 27 cGPS stations of the Campi Flegrei onland monitoring network and the MEDUSA buoys over that period.
At the buoy A site, the same model predicts a horizontal seafloor deformation with a speed $\varv_{M}$ = 3.32 cm/yr and an angle $\alpha_{M} = -112.31^{\circ}$.
From the analysis of the horizontal time series, the authors found instead a speed $\varv_{A} = (5.19 \pm 0.27)$ cm/yr and an angle $\alpha_{A} = (-91.32 \pm 0.89)\,^{\circ}$.
Compared to the model, this measurement presents a significant disagreement, at a level of $\simeq$\,6.9\,$\sigma$,
for the horizontal speed and a large inconsistency, at a level of $\simeq$\,23.7\,$\sigma$, for the deformation angle,
implying a corresponding horizontal seafloor deformation velocity vector pretty incompatible with the prediction
of the Mogi model with the adopted parameters, likely due to the position of this buoy relatively close to the navigation routes in the gulf.

To assess this discrepancy, we developed a new approach including in the analysis both available cGPS and compass data, and applied this method to the buoy A measurements.
The angles provided by the compass sensor are used to analyze the cGPS positions correcting for the effect of the pole inclination,
and finally to derive with a linear best fit the sea bottom horizontal velocity vector at buoy A site.
Indeed, if the effects of environmental forces, which are the main source of the deviations of the pole from the vertical equilibrium position, cannot be assumed as strictly periodic, their effective average contribution does not simply cancel out in the analysis of the time series over a suitable time interval. A proper correction for such deviations is instead necessary to avoid a wrong estimation of the seafloor deformation velocity based on the exploitation of cGPS buoy positions.

Having the CFBA and CFBC buoys the same configuration, we validated the proposed method applying it to the buoy C measurements, but for a limited time interval because of a problem occurred during data acquisition (see App. \ref{buoyC}).

Compared to previous works presented in \cite{xie19} and \cite{demartino20},
the novelty of the methodological approach presented in this paper consists in the combination of four main aspects:

\noindent
$i)$ the estimation of the uncertainties of the compass sensor output angles through a stacking method 
based on the analysis of the real data coming from the cGPS and the compass and its inclusion in the subsequent data analysis;

\noindent
$ii)$ the correction of the cGPS data through compass data, ingesting the previous step in the analysis, 
to account for the pole inclination which, if not taken into account,
can alter the horizontal seafloor deformation estimation;

\noindent
$iii)$ the comparison between the results based on sets of data extracted for different maximum pole inclinations up to 3.5\,$^{\circ}$ and the convergence of the retrieved parameters for increasing pole inclinations;

\noindent
$iv)$ the analysis of the impact of the main systematic effects on the estimated parameters.

The paper is organized as follows. In Sect. \ref{sec:horcor} we describe the theoretical method implemented to derive the horizontal seabed displacement and the associated error jointly accounting for cGPS and compass measurements. Sect. \ref{sec:data} presents the data selection and the analysis, including an original preprocessing aimed at the evaluation of the error to be associated to the compass data and the estimation of the impact of the main systematic effects.
The main results are given in Sect. \ref{sec:res} and compared with the previous ones and with the adopted deformation model. In Sect. \ref{sec:disc} we draw our conclusions and discuss potential developments. In App. \ref{compass} we provide some technical details about the compass sensors and, in App. \ref{buoyC}, we present the results derived applying the same approach to the CFBC data to cross check the validity of the method.

\section{Seafloor horizontal displacement}
\label{sec:horcor}

Let us consider a buoy connected to the seabed with a pole of length $L$. The instantaneous position of the top of the buoy depends on
the orientation of the pole induced by environmental forcing and on possible seafloor motion 
(because of the Campi Flegrei caldera dynamics). 
The amount of the displacement of the buoy top from the vertical position at each sample is given by a set of three angles. 
In the plane defined by the $N$ (South $\rightarrow$ North)  and $E$ (West $\rightarrow$ East) directions, the seabed horizontal displacement at each time series sample is given by the displacement measured by the cGPS and the one recovered from the compass data related to the inclination of the pole. 

The shift due to the pole inclination vanishes when the buoy stands along the vertical position, characterized by zero values of two tilt (roll and pitch) angles. Differently, it can be evaluated on the basis of the rotation matrix defined by the compass data (\cite{xie19}).
Given the pole length, the rotation matrix allows the computation of the two components, ${d}_{E,N}^{C}$, to be subtracted from cGPS horizontal measurements, $d_{E,N}^{G}$, to obtain the corrected seabed horizontal position, $d_{E,N}$.

We choose the ground reference system, solidal with the Earth, consisting of a set of three right-handed axes
along the East ($E$), North ($N$), Up $(U)$ directions (ENU reference system).
Defining the origin of the system in the junction of the pole with the buoy ballast laying on the seabed, when the buoy is unperturbed the cGPS position is determined by the coordinates $(0,0,L)$.
The buoy A, deployed at a depth of 40\,m, has a 43\,m pole connecting the float and subaerial part to the ballast placed at the seabed and the distance between the junction and the cGPS is 47.5\,m, i.e. the $L$ pole length. 
The orientation of the pole
with respect to the ground reference frame can be defined by a set of three Tait-Bryan angles $ \phi,\, \theta,\,\psi$ that
give the rotations around the $E,\, N, \, U$ axis, respectively, according to the right-hand rule.
Here $\phi$ is the pitch angle, $\theta$ the roll angle and $\psi$ the heading angle. 
We set these angles to zero when the pole is in equilibrium, that is
if both its inclination with respect to the vertical direction and its rotation around its own axis, due to a possible twist, vanish. 

We defined $R(\psi,\theta,\phi)$ the rotation that transforms the fixed frame into the frame of the buoy in terms of a sequence of three consecutive rotations around each axis
($R$ is the unitary matrix for $ \phi = \theta = \psi = 0$).
Since rotations do not commute, working in the ENU reference system, we firstly rotate about the $E$ axis, then around the $N$ axis and, finally, about the $U$ axis:
\begin{equation}
R(\psi,\theta,\phi) = R_{U} (\psi) \cdot R_{N} (\theta) \cdot R_{E} (\phi) \, .
\label{eq:rotation}
\end{equation}

The rotation matrix of an angle $\phi$ about the $E$ axis is given by
\begin{equation}
R_{E} (\phi) =
\begin{bmatrix}
 1 & 0 & 0 \\ 
0 & \cos\phi & -\sin\phi \\
0 & \sin\phi & \cos\phi
\end{bmatrix} ,
\label{eq:rot-e}
\end{equation}

\noindent while, the rotation of an angle $\theta$ about the $N$ axis is defined as
\begin{equation}
R_{N} (\theta) =
\begin{bmatrix}
\cos\theta & 0 & \sin\theta \\ 
0 & 1 & 0 \\
-\sin\theta & 0 & \cos\theta
\end{bmatrix} ,
\label{eq:rot-n}
\end{equation}

\noindent and, finally, the rotation of an angle $\psi$ about the $U$ axis is represented by
\begin{equation}
R_{U} (\psi) =
\begin{bmatrix}
\cos\psi & -\sin\psi & 0 \\ 
\sin\psi & \cos\psi & 0 \\
0 & 0 & 1
\end{bmatrix} .
\label{eq:rot-u}
\end{equation}

\begin{figure}
\begin{center}
\includegraphics[width=7cm]{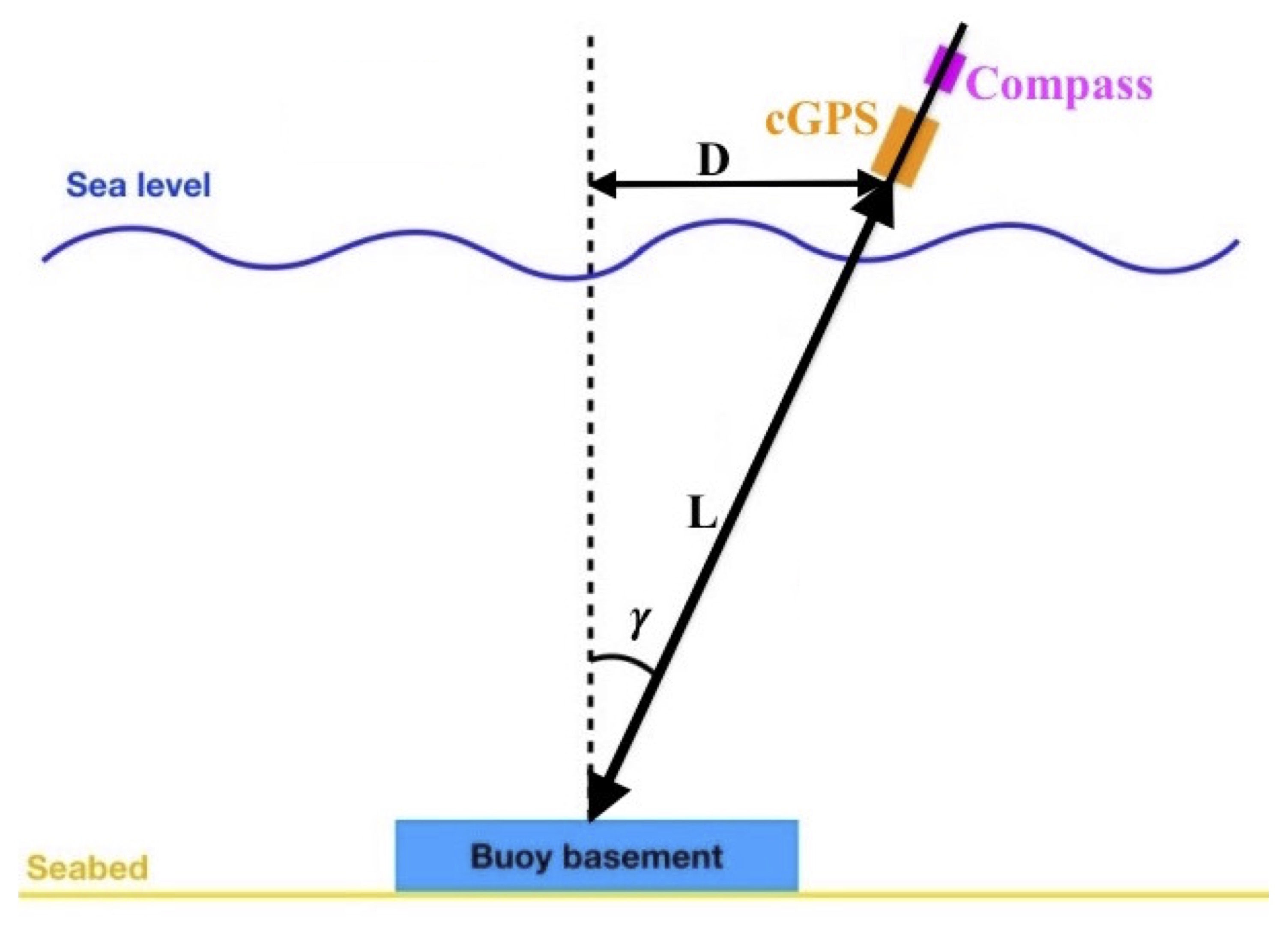}
\end{center}
\caption{Sketch of the buoy. See also the text.}
\label{fig:sketch}
\end{figure}

Applying the rotation matrix to the rotated vector is possible to compute the two horizontal corrections to be subtracted to the cGPS components to theoretically ``bring back'' the buoy in vertical position
\begin{equation}
d_{E}^{C} = L \, [\cos\phi \,\sin \theta\, \cos \psi + \sin\phi\, \sin \psi] \\
\label{eq:corrE}
\end{equation}

\noindent for the $E$ component and
\begin{equation}
d_{N}^{C} =  L \, [\cos\phi \,\sin \theta \,\sin \psi - \sin \phi \,\cos \psi]
\label{eq:corrN}
\end{equation}

\noindent for the $N$ component. $d_{E}^{C}$ and $d_{N}^{C}$ give the distance,
$D = L \, \sin \gamma$
$ = \sqrt{\left({d_{E}^{C}}\right)^2 + \left({d_{N}^{C}}\right)^2}$, 
of the cGPS from the vertical axis passing from the junction of the pole with the ballast, 
$\gamma$ being the angle between the vertical axis and the pole direction, that is the pole inclination;
$\sin^2 \gamma = \cos^2 \phi \, \sin^2 \theta + \sin^2 \phi$, 
with $\gamma \simeq \sqrt{\phi^2 + \theta^2}$ for very small $\phi$ and $\theta$ (see Fig.\,\ref{fig:sketch}).

The propagation error on the correction of the $E$ component is
\begin{align}
\label{eq:errE}
\sigma (d_{E}^{C}) = & \, \Bigl [ \Delta L^{2} \Bigl ( \sin\phi\, \sin\psi + \cos\phi\, \sin\theta \,\cos\psi \Bigr )^{2}
\\ & + L^{2} \Delta \phi^{2} \Bigl ( \cos\phi \,\sin\psi - \sin\phi \,\sin\theta \,\cos\psi \Bigr)^{2}  \nonumber
\\ & + L^{2} \Delta \theta^{2} \Bigl ( \cos\phi \, \cos\theta \,\cos\psi \Bigr )^{2} \nonumber
\\ & + L^{2} \Delta \psi^{2} \Bigl ( \sin\phi \,\cos\psi - \cos\phi \,\sin\theta \,\sin\psi \Bigr)^{2} \Bigr ]^{1/2} \, ,  \nonumber
\end{align}

\noindent
and, analogously, that on the correction of the $N$ component is
\begin{align}
\label{eq:errN}
\sigma (d_{N}^{C}) = & \, \Bigl [ \Delta L^{2} \Bigl ( \sin\phi \,\cos\psi - \cos\phi\, \sin\theta\, \sin\psi \Bigr )^{2}
\\ & + L^{2} \Delta \phi^{2} \Bigl (\cos\phi\, \cos\psi + \sin\phi\, \sin\theta\, \sin\psi \Bigr )^{2}  \nonumber
\\ & + L^{2} \Delta \theta^{2} \Bigl (\cos\phi \, \cos\theta\,\sin\psi  \Bigr)^{2}  \nonumber
\\ & + L^{2} \Delta \psi^{2} \Bigl ( \sin\phi\, \sin\psi + \cos\phi \,\sin\theta \,\cos\psi \Bigr)^{2}\Bigr ]^{1/2} \, .  \nonumber
\end{align}

\noindent In Eqs. \ref{eq:errE} and \ref{eq:errN}, $\Delta L$, $\Delta \phi = \Delta \phi_{C}$, $\Delta \theta = \Delta \theta_{C}$ and $\Delta \psi = \Delta \psi_{C}$ are the rms errors on
$L$, $\phi$, $\theta$ and $\psi$ (or, equivalently, on $\phi_{C}$, $\theta_{C}$ and $\psi_{C}$).

The relationships between $\phi$, $\theta$ and $\psi$ and the output angles provided by the compass, $\phi_{C}$, $\theta_{C}$ and $\psi_{C}$, 
are related to the conventions about the signs of compass outputs. For the compass installed on the buoy A, $\phi = -\phi_{C}$ and $\theta = \theta_{C}$.
Since $\psi$ is defined according to the right-hand rule whereas $\psi_{C}$, as conventional, increases from North to East, they increase in the opposite 
sense. The relationship between $\psi$ and $\psi_{C}$ should also account for: $i$) an offset
between the zero point of the compass and the pole because of 
the arbitrary azimuthal orientation of the compass; $ii)$ the geomagnetic correction, $\psi_{M}(t)$, which is calculated using the World Magnetic Model\footnote{\url{https://www.ngdc.noaa.gov/geomag/WMM/calculators.shtml}} once given the coordinates of the buoy and the period of interest, and turns to be $\psi_{M}(t) \simeq-3.4\,^{\circ}$; $iii)$ the magnetic field induced by the iron buoy pole which affects the measures of $\psi_{C}$, typically of some degrees, 
calling for an accurate calibration of the response curve to properly perform the corresponding correction.
Indeed, the mean of $\psi_{C}$ is very far from zero.
In principle, while the correction term from $i)$ should be a constant, the ones from $ii)$ and $iii)$ depend on the 
detailed distributions of $t$ and $\psi_{C}$, respectively, for the set of samples specifically considered, but the impact of 
the variations of these distributions among the various considered sets of samples in the period of interest is expected be weak. 
In the absence of information about $i$) and $iii$), we can at least remove from the data a suitable estimate of the mean value of $\psi_{C}$ 
to effectively correct for the three effects.
The relationship between $\psi$ and $\psi_{C}$ is then $\psi = - (\psi_{C} - \psi_{0})$, where $\psi_{0}$ is a suitable estimate of the 
mean value of $\psi_{C}$, as discussed in the next sections. 

In general, 
when assuming the buoy in vertical position, $\phi=\theta=0$, the horizontal corrections $d_{E}^{C}$ and $d_{N}^{C}$
vanish (see Eqs. \ref{eq:corrE} and \ref{eq:corrN}), but the various contributions to the associated errors never cancel out and depend on the specific sample
(see Eqs. \ref{eq:errE} and \ref{eq:errN}).
For very small $\phi$, $\theta$ and $\psi$, $d_{E}^{C} \simeq L \, \theta$ and $d_{N}^{C} \simeq -L \, \phi$, and their errors
are dominated by the terms $L \, \Delta \theta$ and $L \, \Delta \phi$, respectively.

\section{Data selection and algorithm}
\label{sec:data}

We considered the data available for the period from 1 July 2017 up to 31 May 2020.
To derive the horizontal displacement of the seafloor, a dedicated routine has been implemented which includes selection criteria and data cleaning to reject the most affected measurements because of bad weather condition and instabilities in the communication system, and the subsequent analysis.

Initially, NaN values due to communication errors between the geodetic cGPS sensors and the ground station have been removed from the records.
Since the cGPS data are sampled every 30\,s while the compass data are recorded each 10\,s, we extracted the only compass data having the same timestamp of the cGPS measurements. Successively, the temporally synchronized data were stored in a single matrix collecting the time, the $E$ and $N$ cGPS values and the three angles from the compass. 
Furthermore, we applied a cleaning routine to discard eventual spikes entering the data by excluding the samples for which at least one of the above variables differs from the average by more than 10 times the corresponding standard deviation, thus avoiding highly contaminated data.

The code allows to perform the analysis in different conditions by introducing a set of cutoff thresholds to be applied to the data. First, we selected data with $U$ positions in a desired interval and with $E$ and $N$ components inside a circular or rectangular area, which are $z_{cut}$ for the $U$ component and $h_{cut}$ for the horizontal directions.
Here, we assumed $z_{cut}$ = $\pm 30$\,cm and a circular selection defined by a radius $h_{cut}$ = 2\,m. We verified that setting the $h_{cut}$ at 4\,m gives substantially the same number of data. The vertical limit, much larger than the measured displacement, further removes data potentially affected by bad weather and tidal conditions.
Then, we defined a compass cutoff variable, $c_{cut}$, equal for the pitch and roll angles to select square boxes.
The $c_{cut}$ parameter was increased of $0.1^{\circ}$ step at each run, starting from the vertical position, which, as already mentioned, corresponds to null pitch and roll angles
($\phi = \theta = 0\,^{\circ}$), up to inclinations identified by $ \mid \phi \mid \, = \, \mid \theta \mid \, = 3.5\,^{\circ} $, for a total of 36 runs.

In general, starting from 2289352 data, 65641 of them ($\sim$ 2.87\,\%) are excluded from the analysis because of missing values, spikes and cGPS cutoff. 
The data analysis assuming the buoy in vertical position does not require any correction to cGPS data (see Eqs. \ref{eq:corrE} and \ref{eq:corrN}), but significantly reduces the fraction  of the considered samples (see Table \ref{tab:ndata}): only a fraction of $\sim 0.59\,\%$ of the remaining $\sim 2.2 \times 10^{6}$ samples satisfies this condition.

Fig. \ref{fig:ndata} represents the data number for each run and the percentage increase, $\Delta I_{\%}$, between two consecutive runs such as, if $i$ and $i+1$ are two successive runs, $\Delta I_{\%, i+1} = (N_{d, i+1}- N_{d, i}) / N_{d, i} \cdot 100$ where $N_{d}$ is the data number. 
As shown in the plot, extending the values of the inclination of the buoy, initially results into a relevant increase of the data accounted in the analysis, but the relative increase of the data number decreases for increasing cutoffs.

Regarding the vertical component, Eq. \ref{eq:rotation} gives a correction

\begin{equation}
d_{U}^{C} =  L \, (1 - \cos\phi \,\cos \theta) \simeq \frac{L}{2} (\phi^{2} + \theta^{2}) \simeq  \frac{L}{2} \gamma^{2}
\label{eq:corrU}
\end{equation}

\noindent to be added to the measured cGPS Up component to consider the pole inclination. We applied a linear fit to the measured vertical shift as function of time accounting or not for the above correction. We verified that including the correction does not change significantly the retrieved best fit velocity, as in principle expected, because the correction depends only at the second order of the pole inclination angle
$\gamma$, differently from the case of the horizontal components where it appears at the first order. 
In principle, the maximum correction value from Eq. \ref{eq:corrU} could achieve $\sim 17$ cm for $\phi = \theta = 3.5^{\circ}$ but we expect that the overall impact on the vertical velocity estimation is strongly suppressed in the fit. Numerically, we find in fact a velocity $\sim 0.7 \,\%$ smaller than the one derived without correction, a difference that amounts to only $1/3$ of the $1\, \sigma$ error quoted in \cite{demartino20}. We find results in excellent agreement with the ones reported in \cite{demartino20}, thus we will focus only on horizontal displacements. 

\begin{table*}
\begin{minipage}{0.22\linewidth}
\centering
\small\addtolength{\tabcolsep}{-4.5pt}
\begin{tabular}{ccc}
\hline  \hline  \\ [-1.5ex]
$\mid \phi \mid = \mid \theta \mid$ & $N_{d}$ & \%\\
(deg) & \\
\hline & \\ [-1.5ex]
0.0 & 13152 & 0.59 \\
0.5 & 778706 & 35.0 \\
1.0 & 1676394 & 75.4\\
1.5 & 1996366 & 89.8 \\
2.0 & 2117465 & 95.2\\
2.5 & 2174474 & 97.8 \\
3.0 & 2204723 & 99.1\\
3.5 & 2222266 & 99.9 \\

\hline \hline & \\ [-1.5ex]
 \end{tabular}
    \caption{Number and percentage of data analyzed at some selected run as function of the maximum value of $\mid \phi \mid$ and $\mid \theta \mid$.}
    \label{tab:ndata}
    \end{minipage}\hfill
\begin{minipage}{0.7\linewidth}
\hskip -0.5cm
\includegraphics[width=9.5cm]{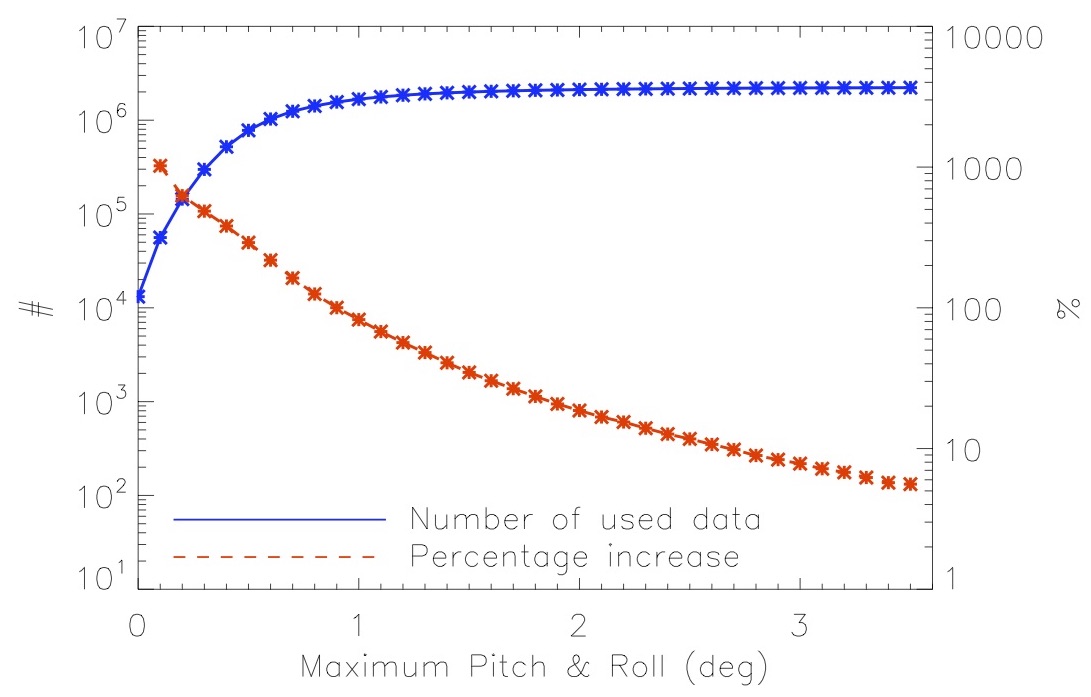}
\captionof{figure}{Number of data for each run (solid blue line and asterisks) and percentage increase with respect to the previous run (dashed red line and asterisks).}
\label{fig:ndata}
\end{minipage}
\end{table*} 

\subsection{Error evaluation from data}
\label{sec:errval}

The errors on the recovered horizontal corrections, given by Eqs. \ref{eq:corrE} and \ref{eq:corrN}, were estimated using the propagation formulas in Eqs. \ref{eq:errE} and \ref{eq:errN}. According to the design drawings and the mechanical construction of the entire pole, a suitable estimation of the rms error on the pole length is $\Delta L = 2$\,cm. The rms errors $\Delta \phi$, $\Delta \theta$ and $\Delta \psi$ in dynamic conditions are expected to be larger than the nominal resolutions of $0.1\,^{\circ}$ and $0.8\,^{\circ}$ (see App. \ref{compass}), representing a suitable estimate of the rms errors of the measurement when taken in static conditions, and need to be quoted from the data, as outlined below.
 
\begin{figure*}[h!]
\begin{center}
\includegraphics[width=7.cm]{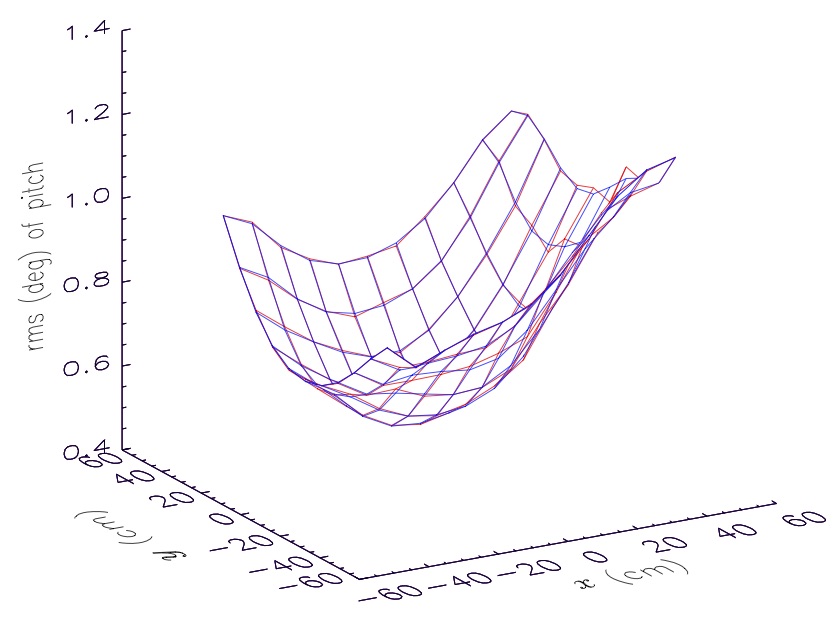}
 \includegraphics[width=7.cm]{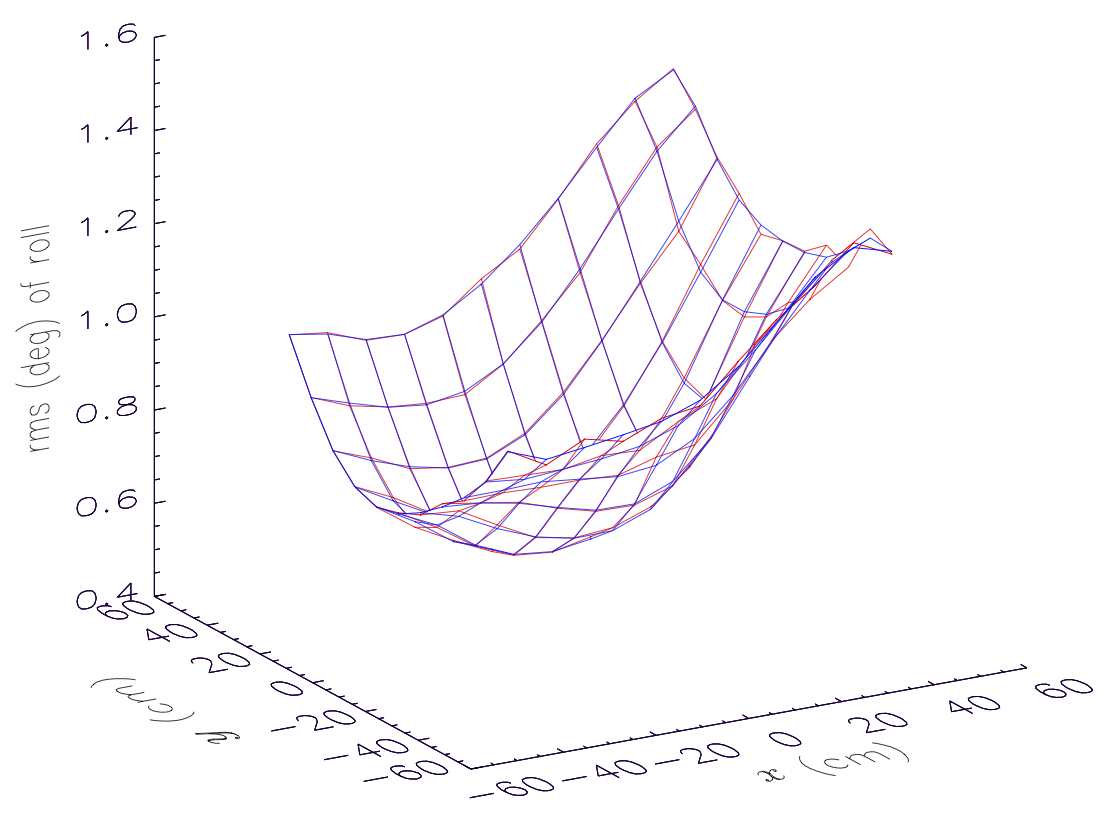}
 \includegraphics[width=7.cm]{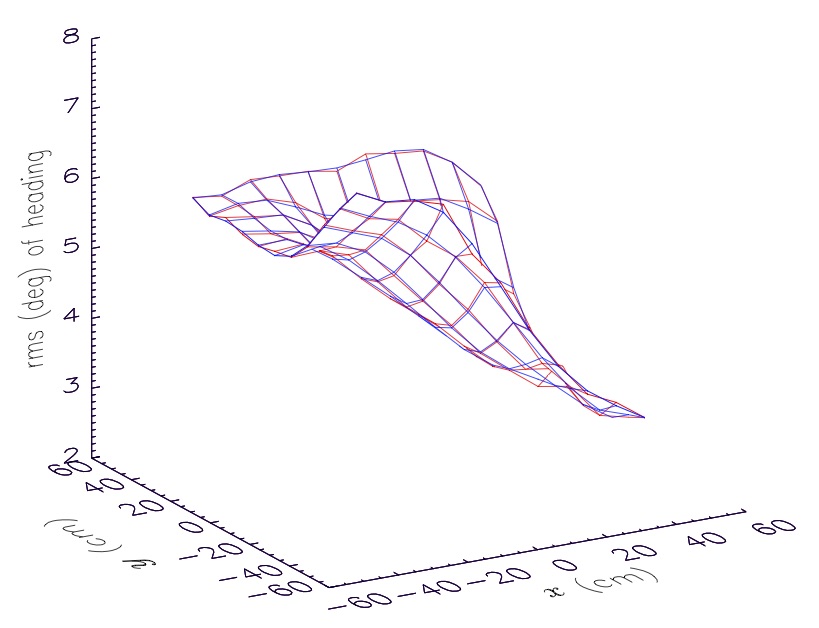}
\end{center}
\caption{Surface representation of the rms uncertainty for the compass output pitch (top panel), roll (middle panel) and heading (bottom panel) angle. In all plots,
we show the stacked rms values (in red) and the corresponding fitted surface (in blue) as function of the displacements $x$ and $y$. The two surfaces are almost superimposed since the fit is extremely accurate.}
\label{fig:surface}
\end{figure*}

We considered data within one week in order to have a time interval small enough to assume that the seabed displacement is negligible, on one side, and large enough to enclose a sufficient number of data for a suitable statistical evaluation, on the other. In this range, we first identified those data having null pitch and roll and computed their averaged values for the $E$ and $N$ components. These averages are then subtracted to all the corresponding data of each chunk to obtain the corresponding displacements,
$x$ and $y$, respectively. 
We then divided the horizontal plane in a fixed grid of $11 \times 11$ points with a step size corresponding to the projection of the 0.1$^{\circ}$ resolution step
of roll and pitch angles, that is $\Delta_{E,N} = L \, \sin(0.1\,^{\circ}) \simeq 8.29$\,cm, 
evaluated the rms of the three compass angles within each week and finally stacked the results of the different weeks. Each of the three resulting surfaces can be well fitted by a sixth order 2D polynomial (see Fig. \ref{fig:surface}) which provides a good estimate of the errors to be associated to $\phi, \, \theta$ and $\psi$. The fitted polynomial function is:
\begin{equation}
f(x, y) = \sum k_{j,i} \cdot x^{i} \cdot y^{j} \, ,
\label{eq:surfit}
\end{equation}

\noindent with $i$ and $j$ from 0 to 6, $k_{j,i}$ being the set of coefficients of a 2D-polynomial function that, for us, corresponds to the pitch, roll and heading rms stacking evaluations
as functions of $x$ and $y$.
In Fig. \ref{fig:surface} we compare the reconstructed surface rms (in red) and the generated fit (in blue) for the three angles. As emerges from the plots, the pitch and roll rms values vary in a small range along the horizontal plane. The behaviour of the heading rms surface, varying from $\simeq$ 3\,$^{\circ}$ up to $\simeq$ 6\,$^{\circ}$, is very different from the other two and it exhibits a stronger variation on the horizontal plane.
Their resulting averaged rms values are $\Delta \phi = (0.800 \pm 0.168)\,^{\circ}$, $\Delta \theta = (0.902 \pm 0.208)\,^{\circ}$ and $\Delta \psi = (4.95 \pm 1.04)\,^{\circ}$ at 1\,$\sigma$ error. 
The polynomial fits represented by Eq. \ref{eq:surfit} well characterize the rms shapes, at least for the 96\,\% of the data, with the module of the relative deviation from the retrieved surface value which is $\lesssim$ 3.5\,$\%$ for the pitch angle and $\lesssim$ 3\,$\%$ for the roll and heading angles. Increasing the degree of the 2D polynomial function does not appreciably improve the quality of the surface representation. For the remaining fraction of data, less than 4\,\%, where the excursion $x$ and $y$ 
from the centre of the grid is far from the region of validity of Eq. \ref{eq:surfit}, we set $\Delta \phi$, $\Delta \theta$ and $\Delta \psi$ to their corresponding averaged rms plus (or minus) 3\,$\sigma$ if the polynomial fit predicts larger (or smaller) values.

To verify the robustness of this error estimates with respect to the size of the data chunk, we finally checked that the above results do not significantly change if we consider a time interval of one month instead of a week.

\subsection{Parameter estimation}
\label{parest}

As anticipated in Sect. \ref{sec:horcor}, we first subtracted a suitable estimate of the compass output heading angle when the buoy is in equilibrium: 
we removed the weighted average 
when the buoy is in vertical position, which gives a value of $\psi_{0} = 37.895\,^{\circ}$.

We performed a $\chi^{2}$ analysis based on the comparison of the expected displacement along the two directions, $d_{E,N}^{fit}$, and the observed quantities
possibly corrected as described above. 

Assuming a linear fit, the model is described by two variables for each of the two components of the horizontal shift $d_{E,N}^{fit} = (A_{E,N} + B_{E,N} \,t)$ as function of time. We performed a fit for each considered cutoff in $\phi$ and $\theta$ characterizing the maximum inclination of the buoy. 
The minimum of the $\chi^{2}$ can be determined minimizing with a least square approach over the two fitting parameters (\cite{Bevington}).
For each run, we evaluated the $\chi^{2}_{E,N}$ along each $E$ and $N$ direction

\begin{equation}
\chi^{2}_{E}  = \sum_{i} \frac{[(A_{E} + B_{E} t_{i}) - d_{E,i}]^{2}}{\sigma^{2} (d_{E,i}^C) + \sigma^{2}_{G}}  \, ,
\label{eq:chi2E}
\end{equation}

\noindent and 
\begin{equation}
\chi^{2}_{N} = \sum_{i} \frac{[(A_{N} + B_{N} t_{i}) - d_{N,i}]^{2}}{\sigma^{2} (d_{N,i}^C) + \sigma^{2}_{G}} \, .
\label{eq:chi2N}
\end{equation}

\noindent where $i=1,\, N_{d}$ and $\sigma_{G} = 3$\,cm is a suitable estimate of the rms error of the cGPs $E$ and $N$ measurements.
In these equations, $d_{E,N} = d_{E,N}^{G} - {d}_{E,N}^{C}$ to perform the correction for the pole inclination
(see also Eqs. \ref{eq:corrE}, \ref{eq:corrN}, \ref{eq:errE} and \ref{eq:errN}).
For the analysis of the raw data, i.e. neglecting the correction for the pole inclination and using only the cGPs data, we obviously set instead
$d_{E,N} = d_{E,N}^{G}$ and $\sigma^{2} (d_{E,N}^C) = 0$. 

The corresponding reduced $\chi^{2}$ are $\chi^{2}_{E,red} = \chi^{2}_{E}/n$ and $\chi^{2}_{N,red} = \chi^{2}_{N}/n$ where $n = N_{d}-2$ is the total number of data (see Table \ref{tab:ndata}) minus the number of fitted parameters.

\section{Results}
\label{sec:res}

In this section, we describe the best fit results on the seafloor displacement derived individually for each adopted cutoff.
The horizontal deformation angle is given by the arctangent of the ratio between the displacement along the $N$ direction and the one along the $E$ direction or, analogously, their corresponding velocities, $\alpha = \text{atan} (d_{N}/d_{E}) = \text{atan} (\varv_{N}/\varv_{E})$.

The error on the speed, $\sigma (\varv)$, can been derived with the standard propagation from the uncertainty in the two velocity components
\begin{equation}
\sigma( \varv) = \left [ \frac{ \varv_{N}^{2} \sigma^{2}(\varv_{N}) + \varv_{E}^{2} \sigma^{2}(\varv_{E})}{\varv_{N}^{2} + \varv_{E}^{2}} \right]^{1/2} \, ,
\label{eq:errv}
\end{equation}

\noindent where the error on the velocity components, $\sigma(\varv_{E,N})$, has been estimated as a weighted mean squared error based on the errors at each time used in the fit procedure evaluated as described in Sects. \ref{sec:horcor} and \ref{sec:errval}. Similarly, the error on the angle, $\sigma(\alpha)$, has been derived through the standard propagation formula which, starting from the definition of $\alpha$, leads to:
\begin{equation}
\sigma(\alpha) = \frac{1}{1+ \left ( \frac{\varv_{N}}{\varv_{E}} \right)^{2}} \left [ \left( \frac{\sigma(\varv_{N})}{\varv_{E}} \right )^{2} + \left ( \frac{\varv_{N}}{\varv^{2}_{E}} \sigma (\varv_{E}) \right)^{2} \right]^{1/2} \,.
\label{eq:erralpha}
\end{equation}

In Table \ref{tab:raw} we report, for some selected cutoffs (first column), the best fit raw horizontal velocity components with their errors
along with the speed, the deformation angle and their corresponding errors calculated, instead, from Eqs. \ref{eq:errv} and \ref{eq:erralpha} and the reduced $\chi^{2}$ summed over the two components. As emerges from the table, the speed is mainly contributed by the $N$ component, as expected from the Mogi model (see Sect. \ref{intro}), but the resulting speed values always significantly exceed the model prediction. The velocity components, particularly for the East one and, correspondingly, the deformation angle are not consistent among the different cutoffs, their differences significantly exceeding their errors. The reduced $\chi^{2}$ are always much larger than unity. Clearly, both the high values of the reduced $\chi^{2}$ and the inconsistency with the model call for a suitable correction accounting for the pole inclination.

\begin{table*}[h!]
\centering
\small\addtolength{\tabcolsep}{-2pt}
{\begin{tabular}{cccccc}
\hline  \hline  \\ [-1.5ex]
$\mid \phi \mid = \mid \theta \mid$ & $\varv_{E}^{raw} \pm \sigma(\varv_{E}^{raw})$ & $\varv_{N}^{raw} \pm \sigma(\varv_{N}^{raw})$ & $\varv^{raw} \pm \sigma(\varv^{raw})$ & $\alpha^{raw} \pm \sigma(\alpha^{raw})$ & $\chi^{2,raw}_{E+N,red}$\\
(deg) & (cm/yr) & (cm/yr) & (cm/yr) & (deg) & \\
\hline & \\ [-1.5ex]
0.0 & -1.657 $\pm$ 0.031 & -5.185 $\pm$ 0.031 & 5.443 $\pm$ 0.031 & -107.728 $\pm$ 0.325 & 61.8\\
0.5 & -1.369 $\pm$ 0.004 & -5.290 $\pm$ 0.004 & 5.465 $\pm$ 0.004 & -104.510 $\pm$ 0.041 & 40.6\\
1.0 & -1.301 $\pm$ 0.003 & -5.414 $\pm$ 0.003 & 5.569 $\pm$ 0.003 & -103.511 $\pm$ 0.028 & 50.5\\
1.5 & -1.362 $\pm$ 0.002 & -5.516 $\pm$ 0.002 & 5.682 $\pm$ 0.002 & -103.867 $\pm$ 0.025 & 63.5\\
2.0 & -1.405 $\pm$ 0.002 & -5.594 $\pm$ 0.002 & 5.768 $\pm$ 0.002 & -104.096 $\pm$ 0.024 & 73.4\\
2.5 & -1.430 $\pm$ 0.002 & -5.667 $\pm$ 0.002 & 5.845 $\pm$ 0.002 & -104.164 $\pm$ 0.023 & 80.4\\
3.0 & -1.439 $\pm$ 0.002 & -5.705 $\pm$ 0.002 & 5.884 $\pm$ 0.002 & -104.159 $\pm$ 0.023 & 85.5\\
3.5 & -1.442 $\pm$ 0.002 & -5.736 $\pm$ 0.002 & 5.914 $\pm$ 0.002 & -104.116 $\pm$ 0.023 & 89.0\\
\hline \hline & \\ [-1.5ex]
    \end{tabular}}
    \caption{Raw data best fit estimation of the velocity components along the horizontal directions, the speed, the deformation angle, their statistical errors and the global reduced $\chi^{2}$ summed over the two components for selected dataset identified by the inclination of the buoy (first column).}
    \label{tab:raw}
\end{table*} 
In Table \ref{tab:corr} we report the same set of quantities recovered from the best fit but after the correction.
As before, the statistical errors quoted accounting for a more limited data sample are correspondingly larger, as expected.

\begin{table*}[h!]
\centering
\small\addtolength{\tabcolsep}{-2pt}
{\begin{tabular}{cccccc}
\hline  \hline  \\ [-1.5ex]
$\mid \phi \mid = \mid \theta \mid$ & $\varv_{E} \pm \sigma(\varv_{E})$ & $\varv_{N} \pm \sigma(\varv_{N})$ & $\varv \pm \sigma(\varv)$ & $\alpha \pm \sigma(\alpha)$ & $\chi^{2}_{E+N,red}$\\
(deg) & (cm/yr) & (cm/yr) & (cm/yr) & (deg) & \\
\hline & \\ [-1.5ex]
0.0 & -1.226 $\pm$ 0.556 & -5.240 $\pm$      0.511 & 5.382 $\pm$      0.513 & -103.168 $\pm$       5.900 & 0.197\\
0.5 & -0.779 $\pm$     0.067 &      -5.333 $\pm$     0.062 &       5.390 $\pm$     0.062 &      -98.306 $\pm$      0.715 &      0.375\\
1.0 & -1.552 $\pm$     0.047 &      -3.667 $\pm$     0.043 &       3.982 $\pm$     0.044 &      -112.940 $\pm$      0.665 &      0.783\\
1.5 & -1.620 $\pm$     0.043 &      -3.203 $\pm$     0.040 &       3.590 $\pm$     0.041 &      -116.829 $\pm$      0.683 &       1.188\\
2.0 & -1.567 $\pm$     0.043 &      -3.132 $\pm$     0.039 &       3.502 $\pm$     0.040 &      -116.580 $\pm$      0.685 &       1.519\\
2.5 & -1.531 $\pm$     0.042 &      -3.191 $\pm$     0.039 &       3.539 $\pm$     0.039 &      -115.631 $\pm$      0.672 &       1.798\\
3.0 & -1.507 $\pm$     0.042 &      -3.178 $\pm$     0.039 &       3.517 $\pm$     0.039 &      -115.377 $\pm$      0.673 &       2.027\\
3.5 & -1.497 $\pm$     0.042 &      -3.187 $\pm$     0.039 &       3.521 $\pm$     0.039 &      -115.159 $\pm$      0.670 &       2.221\\
 \hline \hline & \\ [-1.5ex]
    \end{tabular}}
    \caption{The same as in Table \ref{tab:raw} but derived applying the correction method to account for the pole inclination.}
    \label{tab:corr}
\end{table*} 

It is interesting to compare the results found for the different cutoffs. In the absence of relevant instrumental systematic effects, our technique should provide well compatible results independently of the inclination of the buoy. On the other hand, systematic effects are found to be relevant and cannot be ignored, as discussed in Sect. \ref{syst}.

Increasing the allowed maximum pole inclination and consequently the number of data used in the analysis, we found a good convergence of the results and a better agreement with the Mogi model. 

\begin{table*}[h!]
\centering
\small\addtolength{\tabcolsep}{-2pt}
{\begin{tabular}{ccccc}
\hline  \hline  \\ [-1.5ex]
$\mid \phi \mid = \mid \theta \mid$ & $\varv^{raw} \pm \sigma(\varv^{raw})$ & $\alpha^{raw} \pm \sigma(\alpha^{raw})$ & $\varv \pm \sigma(\varv)$ & $\alpha \pm \sigma(\alpha)$ \\
(deg) & (cm/yr) & (deg) & (cm/yr) & (deg) \\
\hline & \\ [-1.5ex]
0.0 &       5.443 $\pm$      0.168 &      -107.728 $\pm$     1.843 &       5.443 $\pm$     0.168 &      -107.728 $\pm$     1.843 \\
0.5 &       5.465 $\pm$     0.017 &      -104.510 $\pm$      0.187 &       5.532 $\pm$     0.029 &       -99.480 $\pm$      0.295 \\
1.0 &       5.569 $\pm$     0.013 &      -103.511 $\pm$      0.141 &       4.257 $\pm$     0.029 &      -111.775 $\pm$      0.388 \\
1.5 &       5.682 $\pm$     0.014 &      -103.867 $\pm$      0.142 &       3.939 $\pm$     0.033 &      -114.029 $\pm$      0.487 \\
2.0 &       5.768 $\pm$     0.014 &      -104.096 $\pm$      0.146 &       3.875 $\pm$     0.037 &      -113.091 $\pm$      0.557 \\
2.5 &       5.845 $\pm$     0.015 &      -104.164 $\pm$      0.149 &       3.944 $\pm$     0.040 &      -112.094 $\pm$      0.598 \\
3.0 &       5.884 $\pm$     0.015 &      -104.159 $\pm$      0.151 &       3.945 $\pm$     0.042 &      -111.520 $\pm$      0.640 \\
3.5 &       5.914 $\pm$     0.016 &      -104.116 $\pm$      0.152 &       3.965 $\pm$     0.044 &      -111.372 $\pm$      0.669 \\
        \hline \hline & \\ [-1.5ex]
    \end{tabular}}
    \caption{Raw and corrected speed and deformation angle derived replacing the propagation error with the rms of the residuals. See also the text.}
    \label{tab:residual}
\end{table*} 

In Table \ref{tab:residual} we also report the results for the speed, the deformation angle and the corresponding errors using the raw and corrected data with an alternative error treatment. Here, we replace the error propagation analysis described in the previous sections (to quote the error of each data sample) with the rms of the residuals in the data with respect to the corresponding values estimated from the predictions of the best fit.
In this approach, the $\chi^{2}_{red}$ for both directions is unity by construction, thus we omit it in the table.

We note that the errors quoted in the second and third column of Table \ref{tab:residual} are larger than the ones in the corresponding columns of Table \ref{tab:raw}
by about one order of magnitude and a factor around 7 for the speed and the deformation angle, respectively.
Indeed, the error estimation used to derive the results in Table \ref{tab:raw} neglects by construction the effect on the residuals introduced by the pole inclination.
Of course, the best fit values are the same, being based in both cases on the assumption of a constant sample error.

On the contrary, when data are corrected for the effect of pole inclination (fourth and fifth column of Table \ref{tab:residual} and corresponding columns in Table \ref{tab:corr}) the errors are comparable, especially allowing for relatively larger inclinations (cutoff $\ge2\, ^{\circ}$), mutually corroborating the error estimation used in the two types of approaches. Of course, the results 
in Table \ref{tab:residual} are based on the assumption of a constant sample error whereas the error propagation analysis used in Table \ref{tab:corr} allows to consider 
a sample dependent error (other than to perform a non trivial $\chi^{2}$ analysis). It is then not surprising that, in the presence of systematic effects, the two methods provide results that,
although similar, may differ in the details: indeed, neglecting systematic effects, the best fit values derived for the investigated quantities agree only at several $\sigma$ level.
On the other hand, as evident from the analysis reported below (see Sect. \ref{syst}),
they are clearly well compatible once systematic errors, other than statistical ones, are taken into account.

\begin{figure*}[h!]
\begin{center}
\includegraphics[width=11cm]{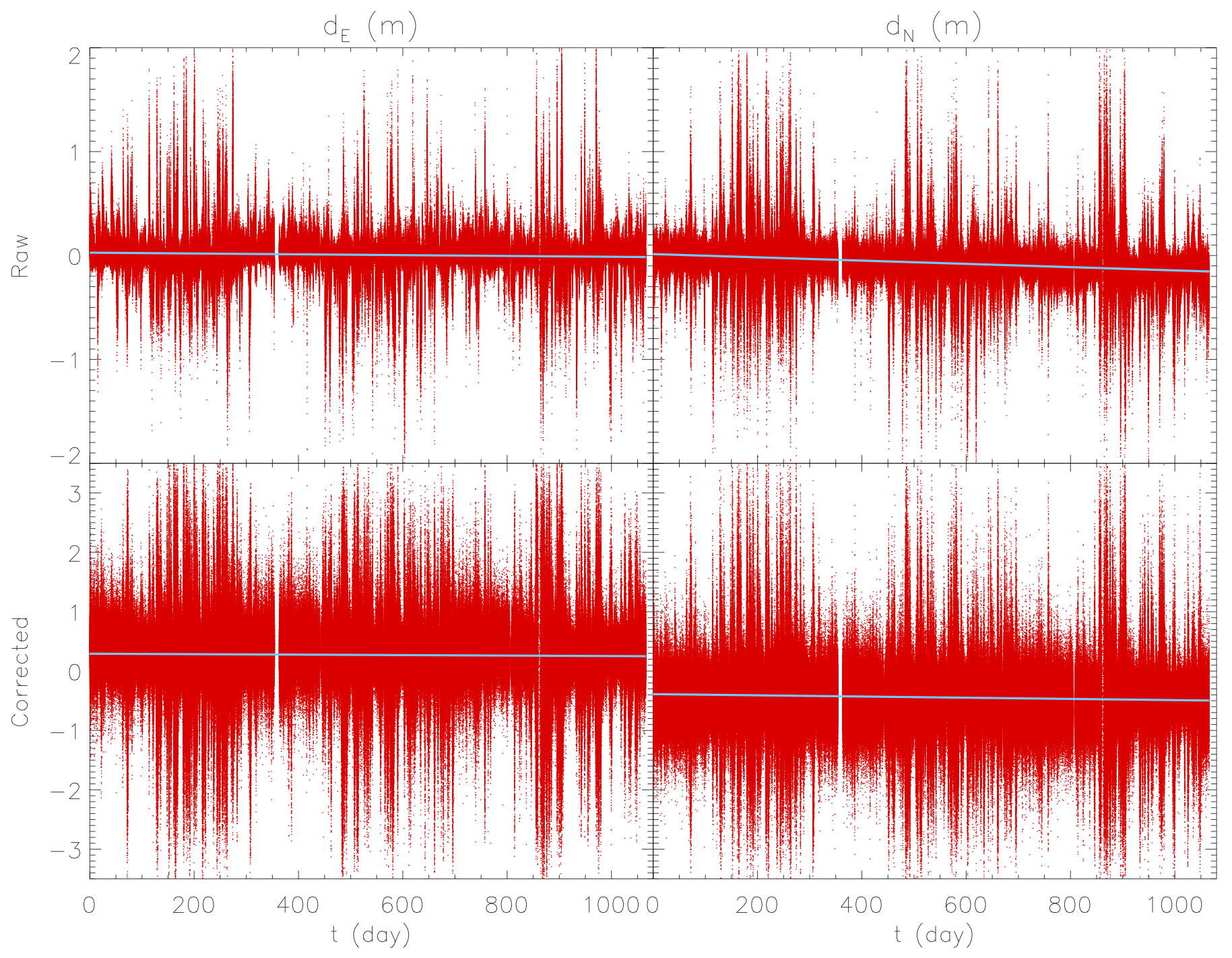}
\end{center}
\caption{Top row: $E$ and $N$ raw components and their relative fit. Bottom row: the same but for the corrected components.
The time here is expressed in days starting from the considered initial date. Data selected with the maximum cutoff accounted in the analysis
($\mid \phi \mid = \mid \theta \mid = 3.5\,^{\circ}$). See also the text.}
\label{fig:rawcorr}
\end{figure*}

\begin{figure*}[h!]
\begin{center}
\includegraphics[width=9cm]{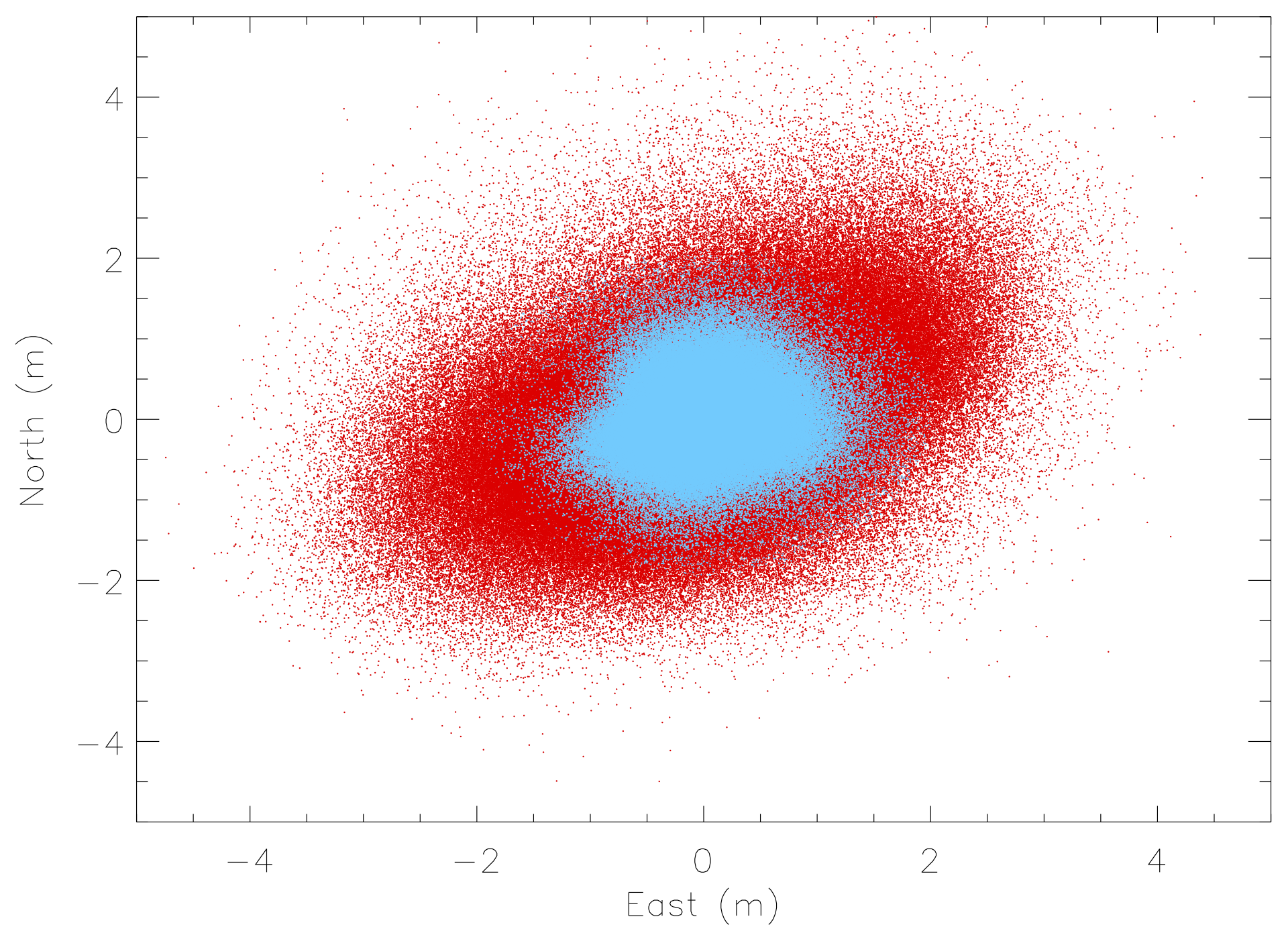}
\end{center}
\caption{$E$ vs $N$ data scattering before (light blue) and after (red) correction: the shape of the distribution is closer to an ellipse after the correction.}
\label{fig:scatter}
\end{figure*}

In Fig. \ref{fig:rawcorr} we show the data for the case relative to the maximum inclination of the buoy before (top row) and after (bottom row) the correction for the two components separately and their corresponding fitting curves. As emerges from the figure, due to the increasing of the overall uncertainty introduced by the compass data, the correction increases the spread of the data that, on the other hand, become more regularly distributed (see Fig.\ref{fig:scatter}) around the best fit curve, alleviating possible bias due to the pole inclination.

\subsection{Systematic effects} \label{syst}

In this section we discuss the impact of the most relevant potential systematic effects.

We first consider the choice of the heading angle characterizing the pole at equilibrium (see Sects. \ref{sec:horcor} and \ref{parest}).
When the pole is in the vertical position, the heading angle has a weighted average of $(37.895 \pm 0.038)\,^{\circ}$ while 
the corresponding average is $(37.443 \pm 7.024)\,^{\circ}$ (rms uncertainties).
Differently,
considering the whole dataset with the cutoff at $3.5\,^{\circ}$, we found a weighted average of $(35.889 \pm 0.003)\,^{\circ}$
and an average of $(35.504 \pm 7.883)\,^{\circ}$.
Thus, globally considering these spreads and differences, we exploit a conservative (minimal) potential offset of the heading measurements, $\delta \psi$, of $10\,^{\circ}$ (2$\,^{\circ}$)
and at each cutoff we assumed a range of $\pm 10\,^{\circ}$ ($\pm 2\,^{\circ}$) to estimate the impact
of a possible instrumental misalignment with the true North in the assumed nominal pole reference system configuration at equilibrium.

Then, we evaluated the impact of a systematic error possibly introduced by a wrong assumption of the pole length of an amount $\delta L$. As a rule of thumb, since we considered a maximum inclination of $ 3.5\,^{\circ}$, the corresponding error on the corrections $d_{E,N}^{C}$ is of the order of $(\theta \,\delta L) \sim 1.22$\,cm for $\delta L \sim 20$\,cm, a somewhat generous upper limit of the error that could come from the assembly of the various components of the pole.
On the other hand, the effective magnitude of the resulting error is expected to be significantly dumped by averaging over a large number of samples. 

Finally, we consider the possible effect of a little inaccuracy in the calibration setup of the compass nominal horizontal plane, estimated under static conditions with the instrumental resolution of $\delta \phi = \delta \theta = 0.1\,^{\circ}$. We then evaluate the systematic errors introduced for the different combinations of offsets $\delta \phi$ and $\delta \theta$ between $-0.1^{\circ}$ and $0.1^{\circ}$.

We compute the errors estimated as the difference between the best fit values obtained assuming the offsets described above, for both negative and positive values of the systematic effect, and the best fit values neglecting the systematics. We evaluated also the half difference of the best fit values obtained for the negative and positive values of the systematic effect. For simplicity and since these three estimations are similar, we conservatively report in Table \ref{tab:systerr} the maximum of them. In the conservative case, the impact of a possible heading offset dominates over the other types of systematic uncertainties, but for small cutoffs. Differently, for the minimal case, the impact of the uncertainty on the pole length is always subdominant, while the other two systematics give comparable impacts.

\begin{table*}[h!]
\centering
\small\addtolength{\tabcolsep}{-2.3pt}
{\begin{tabular}{ccccccccc}
\hline  \hline  \\ [-1.5ex]
$\mid \phi \mid = \mid \theta \mid$ & $(\delta \psi)_{\varv}$ & $(\delta \psi)_{\alpha}$ & $(\delta \psi)_{\varv}$ & $(\delta \psi)_{\alpha}$ & $(\delta L)_{\varv}$ & $(\delta L)_{\alpha}$ & ($\delta \phi = \delta \theta)_{\varv}$ & ($\delta \phi = \delta \theta)_{\alpha}$\\
(deg) & (cm/yr) & (deg)  & (cm/yr) & (deg)  & (cm/yr) & (deg)  & (cm/yr) & (deg)\\
\hline & \\ [-1.5ex]
0.0 & 0.039 & 0.60 & 0.0039 & 0.13 & 8.8 $\times 10^{-6}$ &  5.0 $\times 10^{-5}$ &0.052 & 0.55 \\
0.5 & 0.077 & 0.13 &0.016 & 0.014 & 3.6 $\times 10^{-6}$ & 0.016 & 0.12 & 1.3\\
1.0 & 0.20 &      3.8 & 0.037  &    0.71& 0.0057 &    0.066 & 0.094&       1.3 \\
1.5 & 0.28   &    5.5 & 0.051   &    1.0 & 0.0072  &   0.098 & 0.070 &      1.1\\
2.0 & 0.29    &   6.1 & 0.051    &   1.1& 0.0078   &  0.098 &0.061    &  0.94 \\
2.5 & 0.28     &  6.3 & 0.049     &  1.2& 0.0080   &  0.091 & 0.057    &  0.85\\
3.0 & 0.28     &  6.7 & 0.051      & 1.3& 0.0082   &  0.091 & 0.055     & 0.82 \\
3.5 & 0.29     &  6.8 & 0.052      & 1.3 & 0.0083 &    0.089 & 0.054     & 0.79\\
       
        \hline \hline & \\ [-1.5ex]
    \end{tabular}}
    \caption{Estimate of the errors due to each of the systematic effects discussed in the text for the speed and the deformation angle as specified by the subscript. Second and third columns refer to the conservative heading offset, fourth and fifth to the minimal one.}
    \label{tab:systerr}
\end{table*} 

In the error estimates, considering together these three types of systematic effects other than the statistical errors quoted above, for the maximum allowed pole inclination our method gives a speed $\varv = (3.521 \pm 0.039 \,({stat}) \pm 0.352\, (syst))$ cm/yr and a deformation angle $\alpha = (-115.159 \pm 0.670\, (stat) \pm 7.630\, (syst))\,^{\circ}$,
where the main systematic errors are added linearly and the conservative heading offset case is considered. In the minimal heading offset case, we derive systematic errors of 0.114 cm/yr and 2.150\,$^{\circ}$ for the speed and the angle.

\begin{figure*}
\begin{center}
\includegraphics[width=12cm]{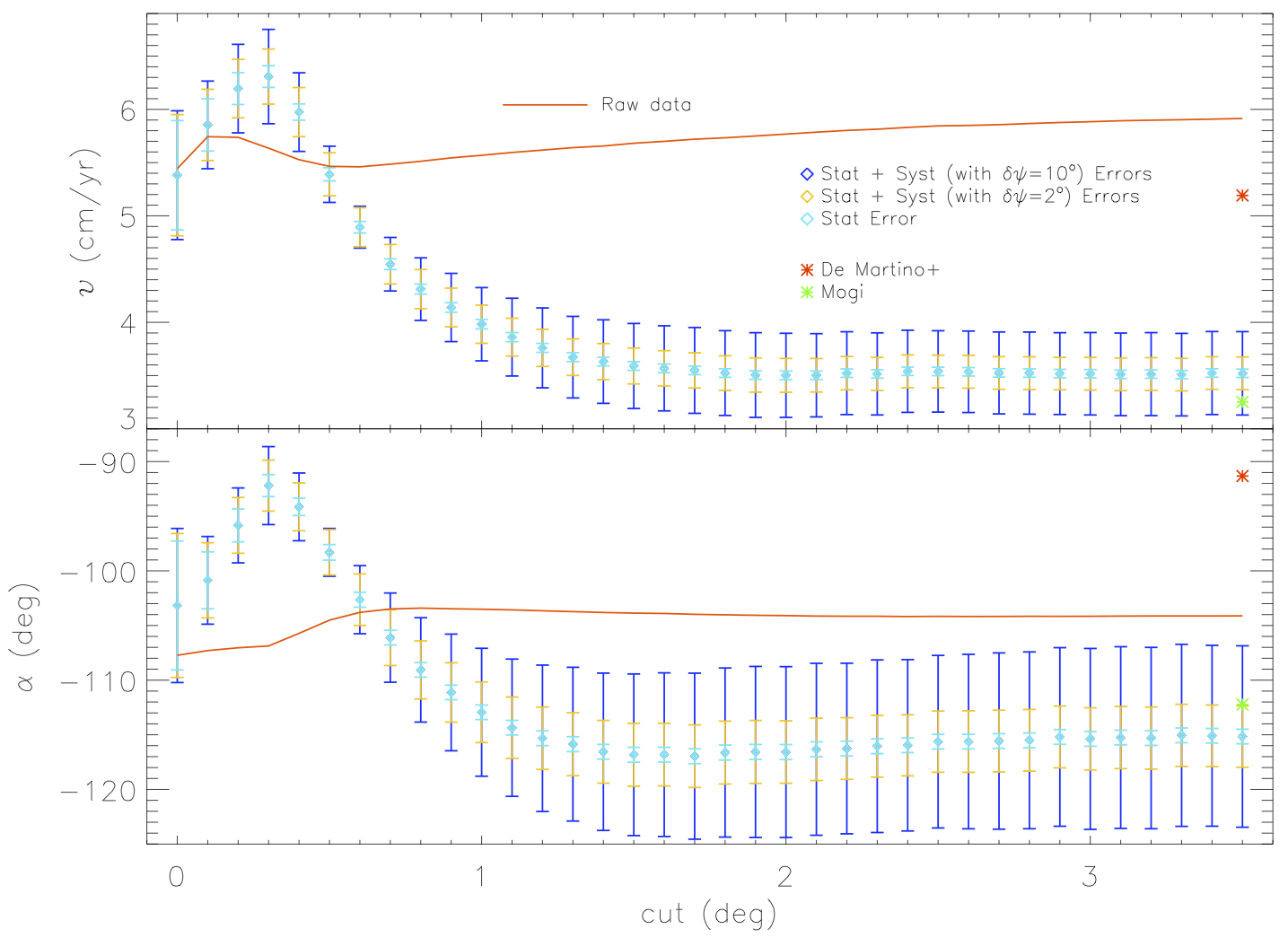}
\end{center}
\caption{Summary of our results applying the correction accounting for pole inclination. The various error bars refers to statistical error alone or to its combination with the systematics for both the conservative and minimal heading offset. For comparison, we also show the results based on the raw data, the one of \cite{demartino20} and the model prediction. See also the text.}
\label{fig:velatan}
\end{figure*}

Fig. \ref{fig:velatan} summarizes our results for the velocity and the deformation angle. 
Globally, passing from the results without the correction for the pole inclination to the corrected ones, the horizontal seafloor deformation speed and angle are significantly closer to the prediction of the Mogi model.

Including the systematics, the speeds and angles retrieved for the various cutoffs are in agreement with each other at $\sim 3\, \sigma$ and $\sim 2\, \sigma$ level, respectively. Accumulating a relatively large fraction of data, i.e. from a cutoff $\gsim\,0.7\,^{\circ}$, the compatibility further improves. 

\begin{figure}[h!]
\begin{center}
\includegraphics[width=8.8cm]{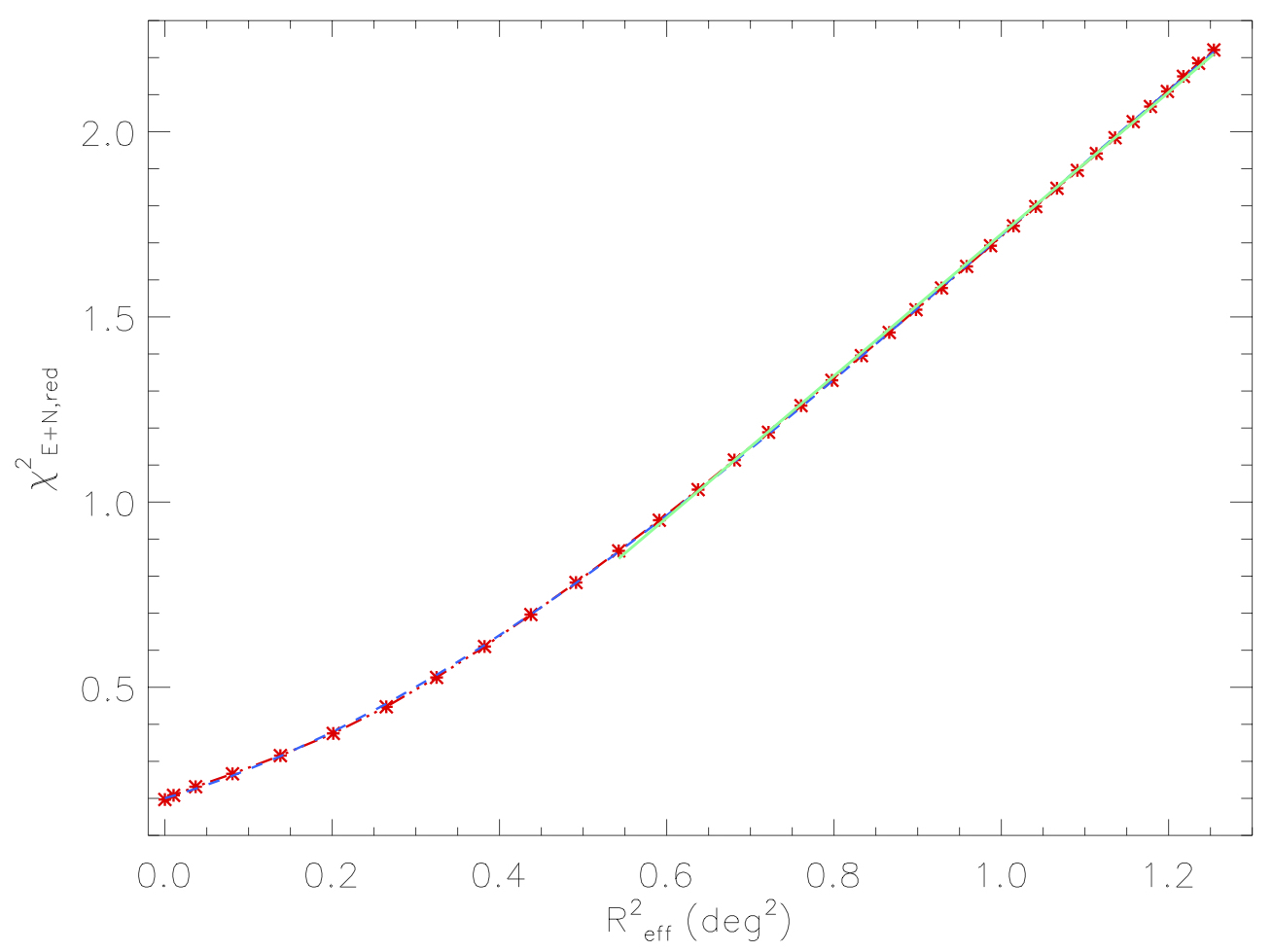}
\end{center}
\caption{Dependence of $\chi^{2}_{E+N,red}$ on $R^{2}_{\rm eff}$. Red dash-dotted line and asterisks refer to the real data; blue dashed line refers to the corresponding third order polynomial fitting; green solid line refers to a linear fit for $R^{2}_{\rm eff} \gsim 0.54$. See also the text.}
\label{fig:cutntrue}
\end{figure}

It is interesting to note that the values of $\chi^{2}_{E+N,red}$ in Table \ref{tab:corr} increase with the cutoff. For each cutoff, we evaluated the effective square of the pole inclination, $R^{2}_{\rm eff} \simeq \left < (\theta^{2} + \phi^{2}) \right >$, averaged over the corresponding samples. As shown in Fig. \ref{fig:cutntrue}, while the $\chi^{2}_{E+N,red} (R^{2}_{\rm eff})$ function is well characterized by a third degree polynomial, it exhibits an almost linear dependence for sufficiently large values of $R^{2}_{\rm eff}$. This linear behaviour has a simple explanation considering that the pitch and roll measurements are taken by an accelerometer working mainly in dynamical conditions. 

For instance, if roll and pitch measurements are acquired when the buoy turns into its equilibrium position, the compass accelerometer records a further horizontal acceleration proportional to the amount of deflection from the vertical, which results from the difference between the (buoyancy) force applied to the centre of buoyancy and the gravitational force applied in the centre of mass of the buoy. Hence, including samples with not negligible pitch and roll angle values, a systematic effect arises, which is not taken into account in our method: the amplitude of this effect is proportional to the inclination angle, $\gamma$ (see Fig. \ref{fig:sketch}). In principle, it could be possible to verify if a given data sample is acquired under dynamic or static conditions through a comparison among contiguous data and to apply a proper correction, but this is beyond the scope of this work.
So, assuming that the measurement error, $\varepsilon$, on these angles comes from two contributions, one static and one dynamic, $\varepsilon_{s}$ and $\varepsilon_{d}$, 
neglecting their correlation and other effects, and calling $\varepsilon_{q}$ the one used in the $\chi^{2} $ analysis, we have that $\chi^{2}_{E+N,red} \propto \sum_{i} (\varepsilon_{s,i}^{2} + \varepsilon_{d,i}^{2})/\varepsilon_{q,i}^{2}$. For an accelerometer, it is reasonable to assume that $\varepsilon_{d}$ is proportional to the acceleration, $a$, of the buoy, $\varepsilon_{d} \propto a \propto \gamma$
with $\gamma \simeq (\phi^{2} + \theta^{2})^{1/2} \ll 1$. On average, this leads to a simple approximate relation like $\chi^{2}_{E+N,red} (R^{2}_{\rm eff}) \sim k + m \, R^{2}_{\rm eff}$, where $k$ and $m$ are two constants, which is indeed the dependence shown in Fig. \ref{fig:cutntrue}:
a linear fit to $\chi^{2}_{E+N,red} (R^{2}_{\rm eff})$ well describes our data, the module of the relative difference between the fit and the curve retrieved from data being always less
than $\simeq 2.4\%$ for $R^{2}_{\rm eff} \gsim 0.54$, and typically much smaller.

For a comparison with the results presented in \cite{demartino20},
we considered as combined error parameter, $\sigma_{c}$, the sum in quadrature of the errors of the two analyses that
for the speed is $\sigma_{c}(\varv) = ( \sigma^{2}(\varv_{A}) + \sigma^{2}(\varv) )^{1/2} = 0.27$ cm/yr and, analogously,
for the angle is $\sigma_{c}(\alpha) = ( \sigma^{2}(\alpha_{A}) + \sigma^{2}(\alpha))^{1/2} = 1.11 \,^{\circ}$,
if we consider only the statistical error of our method:
the compatibility is only at a level of $\simeq$\,6.2 $\sigma_{c}$ and of $\simeq$\,21.5 $\sigma_{c}$ for the speed and the deformation angle, respectively.
Adding linearly the conservative systematical and statistical errors of our method, the compatibility turns to be at a level of $\simeq$\,3.6 $\sigma_{c}$ and of $\simeq$\,2.9 $\sigma_{c}$ for the speed and the deformation angle, respectively. 

Compared to the previous work, our analysis solves the discrepancy with the adopted Mogi model. The consistency turns to be at $\simeq$\,5.2\,$\sigma$ ($stat$ only) or $\simeq$\,0.5\,$\sigma$ ($stat$ and $syst$ added linearly) level for the horizontal speed\footnote{Replacing the full error propagation analysis with the rms of the residuals, this discrepancy worses to $\simeq 1.6\,\sigma$ ($stat$ + $syst$).} and, remarkably, at $\simeq$\,4.3\,$\sigma$ ($stat$ only) or $\simeq$\,0.3\,$\sigma$ ($stat$ and $syst$ added linearly) level for the deformation angle. Considering the error induced by the minimal heading offset, we find, instead, a compatibility with the model at $\simeq$\,1.3\,$\sigma$ for the speed and $\simeq$\,1\,$\sigma$ for the angle.

We can conclude that the new implemented method well reconstructs the investigated quantities, significantly improving the results obtained with data selection only without applying any correction to the data to account for the pole inclination. As clearly shown in Fig. \ref{fig:velatan}, our analysis gives a good agreement with the horizontal seafloor deformation velocity predicted by the reference Mogi model.

\section{Conclusion and discussion}
\label{sec:disc}

We devised a new approach to characterize the horizontal seafloor displacement by jointly taking into account the data from cGPS and compass instruments mounted on top of two of the four buoys in the Gulf of Pozzuoli.
In this study, we assumed the Mogi model, which is based on hydrostatic pressure change for a spherical pressure point source in an elastic half-space, with
parameters derived from cGPS Campi Flegrei onland monitoring network and MEDUSA buoys measurements as the reference model to which compare our results.
We focused our analysis to the buoy A data since, in a previous work, the corresponding recovered parameters were well far from the assumed reference model, while buoys B and C gave compatible results. Nevertheless, having the buoy A the same configuration of the buoy C, we also verified the validity of the method for this buoy but in a limited time interval, where compass data are likely not corrupted.
The method investigates 36 different maximum inclinations of the buoys, up to maximum pitch and roll angles of $3.5^\circ$, 
and derives the speed and the deformation angle of the horizontal displacement for the corresponding data samples. 

Being unknown the exact orientation of the compass with respect to a fixed reference frame and the influence of the magnetic field induced by the iron buoy pole on the heading angle measurements, we characterized the equilibrium configuration of the pole reference system considering the condition of vanishing inclination and the weighted average, $\psi_{0}$, of the heading angle output, taken in this condition, as a quantity to be subtracted from the corresponding data samples.
After a proper selection of the data and the estimation of the errors to be associated to the compass angles directly based on the real measurements through a stacking procedure,
we implemented a $\chi^{2}$ analysis to retrieve the parameters of interest for the considered configurations. The reduced $\chi^{2}$ are around unity and
the results found for the different configurations are internally compatible within the errors when including systematic effects. The resulting speed and deformation angle 
clearly converge for increasing maximum allowed cutoffs of the tilt angles.
We evaluated the impact of the main sources of systematic effects: for very small cutoffs it is smaller than the (obviously relatively large) statistical uncertainty, whereas at cutoffs half a degree higher it dominates over the statistical error.

\begin{figure*}[h!]
\begin{center}
\includegraphics[width=12 cm]{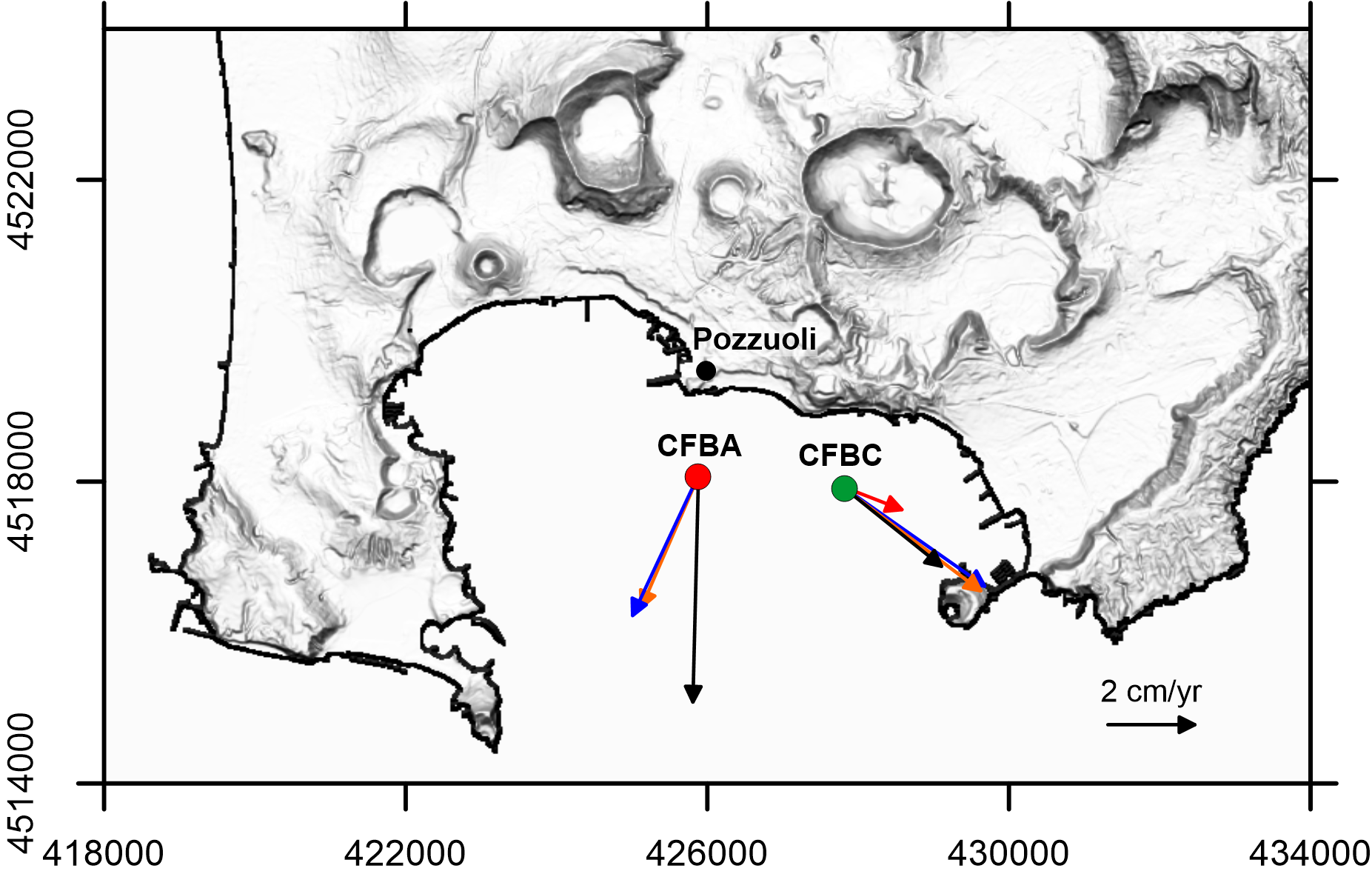}
\end{center}
\caption{Map of CFBA and CFBC cGPS stations (red and green circle) operating in the Campi Flegrei area. The arrows indicate the horizontal velocities found in this study (blue), derived from the Mogi model (orange) and the result from \cite{demartino20} (black). For CFBC buoy we also display (red) the result found applying the method of \cite{demartino20} to the limited period considered in App. \ref{buoyC}.} 
\label{fig:mappa}
\end{figure*}

Fig. \ref{fig:mappa} represents the sea bottom displacement at buoy A and C locations in the map of the Pozzuoli caldera. The three arrows show our results (blue) in comparison with the Mogi model (orange) and with \cite{demartino20} (black). The method developed in this work gives horizontal seafloor deformation velocity vectors in good agreement with the predictions of the adopted Mogi model.
For the buoy A, the module of the vectorial difference between the velocity retrieved from the data and the velocity of the adopted Mogi model changes from 
$ \mid \vec{\varv}_A - \vec{\varv}_M \mid \,= (2.449 \pm 0.241)$\,cm/yr in the case of \cite{demartino20} to
$\mid \vec{\varv} - \vec{\varv}_M \mid\, = (0.320 \pm\, 0.041 \,(stat) \,{\rm or} \,\pm 0.240 \,(stat+syst))$\,cm/yr 
assuming a conservative heading offset (for simplicity, we propagate here the systematic error on the velocity module fixing the angle $\alpha$ to its best fit), then decreasing by a factor $f \simeq 7.65 \pm 1.23 \,(stat) \,{\rm or} \,\pm 5.78 \,(stat+syst)$. Instead, assuming a minimal heading offset, we obtain combined errors of 0.094 cm/yr and 2.37 for $\mid \vec{\varv} - \vec{\varv}_M \mid$ and $f$.
	
The internal consistency, the convergence and the improved agreement with the reference model found in our results support the validity of the analysis,
in spite of the instrumental uncertainties affecting the data. Indeed, we underlined that these results have been achieved despite the compass installed on the buoy was not designed for this purpose, but just for monitoring the buoy status. The developed method applied to the MEDUSA buoys allows the assessment with geodetic accuracy of the seafloor, up to about one hundred meters deep, enabling the extension of a cGPS monitoring network to the whole continental shelf.
The achieved accuracy allows also the kinematic description of the seafloor deformation of a given area by making use of a suitable number of buoys.

We believe that a revised experimental setup and calibration strategy of the acquiring system, specifically designed for studying the horizontal seabed displacement, would allow to further improve the reconstruction of the key parameters, reducing, in particular, the uncertainty due to the systematic effects.
A list of potential improvements includes: $i$) the increasing of the sampling rate and the synchronization of the two instruments; $ii$) a better calibration of the heading angle
provided by the compass in real conditions or the use of a set of gyroscopes for a more precise measurement of the three Tait-Bryan angles; $iii$) the installation of current meters and thermometers along the buoy pole; $iv)$ the development of a more complete model to process the data, able to take into account the effect of the wind and marine current forcing and the mechanical response of the buoy structure, to optimize the scientific result achievable with the above improvements.

\vskip 0.5cm

\backmatter

\bmhead {Acknowledgments}

\noindent This work has been performed in the context of the "Protocollo di intesa di collaborazione scientifica" between INAF-IRA and INGV 
AOO:INAF IRA - Classif. III/10 Prot. n. 0000485, 23 March 2021. 
It is a pleasure to thank the anonymous referees for comments that helped improve the paper.

\begin{appendices}

\section{} \label{compass}

The high-accuracy 3D digital compass sensor consists in an electronic device (with enclosure) using USA patented technology of hard and soft magnetic calibration algorithm; this allows the compass to suppress the magnetic influence through a calibration algorithm in the magnetic interference environment.

The digital compass\footnote{DCM260B-232 datasheet from Rion Technology co., Ver. 1.2, 2020} integrates three-axis fluxgate sensor in real-time solver heading through an internal central processor unit. The three-axis accelerometer is used to perform heading compensation for the possible wide range of tilt angles. This ensures high-precision heading data for tilt angles up to $\pm$\,85$^{\circ}$. Electronic compass integrates high-precision MCU control with output on a standard serial interfaces RS232 and a maximum data-rate of 20 Hz.

During the long-time measurement period (2017-2021), the compass tilt angles recorded are all within of $\pm$10$^{\circ}$, and mostly $\pm$2$^{\circ}$ for both CFBA and CFBC buoys. For these data values, the nominal repeatability for heading measurements should be best of $\pm$0.8$^{\circ}$, while pitch and roll measurements should be best of $\pm$0.1$^{\circ}$.

While pitch and roll measurements are controlled by the force of gravity on fluxgate in the tilt sensor, heading measurements represent changes of buoys orientations and are sensitive to the local magnetic environment and its changes. Therefore, variation of local magnetic environment (e.g., caused by orientation of the iron over- and under-structures of both buoys) can induce offset in the heading measurements. Thus, heading output from the digital compass may be offset from the true heading direction. Leaving this offset uncorrected would impart a significant systematic error to the tilt correction, affecting the precision and accuracy of the horizontal movements estimates for the ballast. In order to avoid this bias, we calibrated each sensor directly in the end-use environment (both buoys), using RION 3D debugging software.

The pitch and roll measurements should not have significant offsets since they use a gravitational reference.

\section{} \label{buoyC}

\begin{figure*}
\begin{center}
\includegraphics[width=12cm]{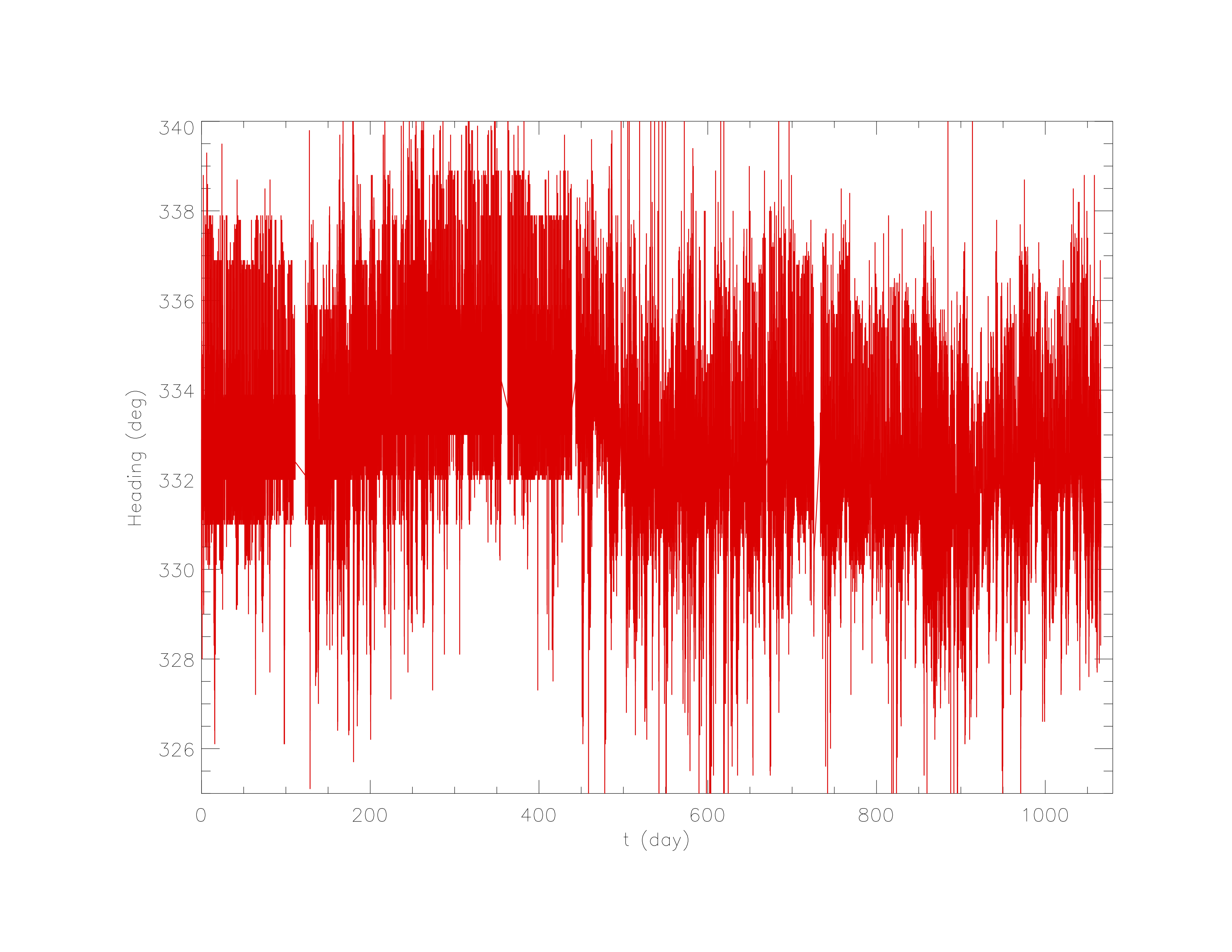}
\end{center}
\caption{Heading angle time series for the buoy C.}
\label{fig:headc}
\end{figure*}

To validate the described method, we applied the same analysis to the CFBC data, having this buoy the same configuration of the buoy A. Unfortunately, one of the two chains of the C buoy, connecting the pole to the ballast, aimed at reducing the pole rotation around its axis, due to possible twists, and at improving its stability, experienced a rupture. Furthermore, we cannot be confident about the overall good quality of the acquired compass data, especially about the heading angle. Indeed, as emerges from Fig. \ref{fig:headc}, where we show the heading data, there is a clear evidence of a different behaviour in the data distribution before and after the $ \sim 430^{\rm {th}}$ day, corresponding to a date around Thursday 13 September 2018. For this reason, we decided to limit our analysis to the data before this date. The Mogi model gives a speed $\varv_{M} = 3.9$ cm/yr and a deformation angle $\alpha_{M} = -36.87\,^{\circ}$. Even though the results of our method at the different cutoffs of the tilt angles are not so stable, especially for the deformation angle, the reliability of the method should improve for increasing data number and, actually, the agreement with the model turns to be very good for the maximum allowed pole inclination. In spite of the retrieved $\chi^{2}_{E+N,red}$ = 3.373, larger than in the case of the buoy A, we find $\varv$ = (3.909 $\pm$ 0.167) cm/yr and $\alpha = (-34.968 \pm 2.507)^{\circ}$ (statistical errors), in excellent agreement with the model. This improves the result $\varv$ = (2.816 $\pm$ 0.266) cm/yr and $\alpha = (-38.360 \pm 5.060)^{\circ}$ (statistical errors) of \cite{demartino20}, in particular for the speed (see Fig. \ref{fig:mappa}). Furthermore, their method applied to the same period considered in this appendix gives $\varv$ = (1.410 $\pm$ 0.455) cm/yr and $\alpha = (-20.336 \pm 14.492)^{\circ}$ (statistical errors), a result which is inconsistent with the Mogi model. The statistical uncertainties found for the two horizontal velocity components scale approximately as the inverse square root of the time length of the data. Thus, considering such a limited time interval, the inconsistency of the retrieved values of $\varv$ and $\alpha$ with respect to the Mogi model does not change significantly in terms of $\sigma$ level, whereas their absolute differences from the model values markedly increase.
The chance to suppress the effective average contribution of environmental forces should reduce with the decrease of the time length of the data, further highlighting the relevance of the method presented in this work.

\end{appendices}

\section*{Declarations}

\begin{itemize}

\item Funding -- TT gratefully acknowledge financial support from the research program RITMARE SP3 WP3 AZ3 U02 and the research contract SMO at CNR/ISMAR
and from the INGV grant Notice n. 4/2021, Prot. n. 6591, 14.04.2021, funded by the EMSO MIUR project (OB.FU. 0530.010).

\item Conflict of Competing interests -- The authors declare that they have no known competing financial interests or personal relationships that could have appeared to influence the work reported in this paper.

\item Ethics approval -- Not applicable
\item Consent to participate -- Not applicable
\item Consent for publication -- Not applicable
\item Availability of data and materials -- As in De Martino et al. 2020 paper, the cGPS data used are available online at \url{http://portale.ov.ingv.it/medusa/download/four-years-of-continuous-seafloor-displacement-measurements}. Compass data are available under request to the Authors.

\item Code availability -- Not applicable

\item Authors' contributions -- Conceptualization, Methodology, Formal Analysis and Validation: Tiziana Trombetti, Carlo Burigana, Francesco Chierici; Data curation: Tiziana Trombetti; Software, Visualization: Tiziana Trombetti, Carlo Burigana; Investigation and Resources: Prospero De Martino, Sergio Guardato, Giovanni Macedonio, Giovanni Iannaccone; Writing -- original draft: Tiziana Trombetti, Carlo Burigana; Writing -- review \& editing: Tiziana Trombetti, Carlo Burigana, Prospero De Martino, Sergio Guardato, Giovanni Macedonio, Giovanni Iannaccone, Francesco Chierici.

\end{itemize}


\begin{thebibliography}{12}
\providecommand{\natexlab}[1]{#1}
\providecommand{\url}[1]{{#1}}
\providecommand{\urlprefix}{URL }
\providecommand{\doi}[1]{\url{https://doi.org/#1}}
\providecommand{\eprint}[2][]{\url{#2}}
 \bibcommenthead

\bibitem[{Bevington and Robinson(2003)}]{Bevington}
Bevington PR, Robinson DK (2003) Data reduction and error analysis for the
  physical sciences, 3rd edn. McGraw Hill, New York, NY

\bibitem[{De~Martino et~al.(2014)De~Martino, Guardato, Tammaro,
  Vassallo, and Iannaccone}]{demartino14b}
De~Martino P, Guardato S, Tammaro U, et~al (2014) A first {GPS}
  measurement of vertical seafloor displacement in the {Campi} {Flegrei}
  caldera ({Italy}). J Volcanol Geotherm Res, 276:145--151.
  \doi{10.1016/j.jvolgeores.2014.03.003},
  \urlprefix\url{https://www.sciencedirect.com/science/article/pii/S037702731400081X}
  
\bibitem[{De~Martino et~al.(2020)De~Martino, Guardato, Donnarumma, Dolce,
  Trombetti, Chierici, Macedonio, Beranzoli, and Iannaccone}]{demartino20}
De~Martino P, Guardato S, Donnarumma GP, et~al (2020) Four years of continuous
  seafloor displacement measurements in the {Campi} {Flegrei} caldera. Front
  Earth Sci 8:641. \doi{10.3389/feart.2020.615178},
  \urlprefix\url{https://www.frontiersin.org/article/10.3389/feart.2020.615178}

\bibitem[{Henderson(1977)}]{Henderson77}
Henderson DM (1977) Euler angles, quaternions, and transformation matrices for
  space shuttle analysis. Design note 1.4-8-020, NASA Astronautics Division,
  Houston,
  \urlprefix\url{https://ntrs.nasa.gov/archive/nasa/casi.ntrs.nasa.gov/19770019231.pdf}

\bibitem[{Iannaccone et~al.(2009)Iannaccone, Guardato, Vassallo, Elia, and
  Beranzoli}]{Iannaccone09}
Iannaccone G, Guardato S, Vassallo M, et~al (2009) A new multidisciplinary
  marine monitoring system for the surveillance of volcanic and seismic areas.
  Seismol Res Lett 80(2):203--213. \doi{10.1785/gssrl.80.2.203},
  \urlprefix\url{https://doi.org/10.1785/gssrl.80.2.203}

\bibitem[{Iannaccone et~al.(2010)Iannaccone, Vassallo, Elia, Guardato, Stabile,
  Satriano, and Beranzoli}]{Iannaccone10}
Iannaccone G, Vassallo M, Elia L, et~al (2010) {Long-term Seafloor Experiment
  with the CUMAS Module: Performance, Noise Analysis of Geophysical Signals,
  and Suggestions about the Design of a Permanent Network}. Seismol Res Lett
  81(6):916--927. \doi{10.1785/gssrl.81.6.916},
  \urlprefix\url{https://doi.org/10.1785/gssrl.81.6.916}

\bibitem[{Iannaccone et~al.(2018)Iannaccone, Guardato, Donnarumma, De~Martino,
  Dolce, Macedonio, Chierici, and Beranzoli}]{Iannaccone18}
Iannaccone G, Guardato S, Donnarumma GP, et~al (2018) Measurement of seafloor
  deformation in the marine sector of the {Campi} {Flegrei} caldera ({Italy}).
  J Geophys Res Solid Earth 123(1):66--83. \doi{10.1002/2017JB014852},
  \urlprefix\url{https://agupubs.onlinelibrary.wiley.com/doi/abs/10.1002/2017JB014852}

\bibitem[{Masterlark(2007)}]{Masterlark}
Masterlark T (2007) Magma intrusion and deformation predictions: Sensitivities
  to the {Mogi} assumptions. J Geophys Res Solid Earth 112(B6).
  \doi{10.1029/2006JB004860},
  \urlprefix\url{https://agupubs.onlinelibrary.wiley.com/doi/abs/10.1029/2006JB004860}

\bibitem[{Mogi(1958)}]{mogi1958relations}
Mogi K (1958) Relations between the eruptions of various volcanoes and the
  deformation of the ground surfaces around them. Bull Earthq Res Inst Univ
  Tokyo 36:99--134

\bibitem[{Sarsito et~al.(2019)Sarsito, Kriswati, Meilano, Andreas, and
  Pradipta}]{Sarsito_2019}
Sarsito DA, Kriswati E, Meilano I, et~al (2019) Volcano deformation monitoring
  using geodetic method: optimal network design. {IOP} Conference Series: Earth
  and Environmental Science 389(1):012039.
  \doi{10.1088/1755-1315/389/1/012039},
  \urlprefix\url{https://doi.org/10.1088/1755-1315/389/1/012039}

\bibitem[{Xie et~al.(2019)Xie, Law, Russell, Dixon, Lembke, Malservisi, Rodgers,
  Iannaccone, Guardato, Naar, Calore, Fraticelli, Brizzolara, Gray, Hommeyer,
  and Chen}]{xie19}
Xie S, Law J, Russell R, et~al (2019) Seafloor geodesy in shallow water with
  {GPS} on an anchored spar buoy. J Geophys Res Solid Earth
  124(11):12116--12140. \doi{10.1029/2019JB018242},
  \urlprefix\url{https://agupubs.onlinelibrary.wiley.com/doi/abs/10.1029/2019JB018242}

\end{thebibliography}
\end{document}